\documentclass[fleqn,usenatbib]{mnras}

\usepackage{newtxtext,newtxmath}
\usepackage[T1]{fontenc}
\usepackage{ae,aecompl}

\usepackage{multirow}
\usepackage{threeparttable}
\usepackage{subfig}
\usepackage{xcolor,colortbl}   
\usepackage{bm} 
\usepackage{graphicx}	
\usepackage{amsmath}	
\usepackage{amssymb}	

\newcommand{\matr}[1]{\bm{#1}} 

\definecolor{lightgray}{rgb}{0.83, 0.83, 0.83}

\title[C-BASS: Simulated parametric fitting in single pixels]{The C-Band All-Sky Survey (C-BASS): Simulated parametric fitting in single pixels in total intensity and polarization}

\author[L. Jew et al.]{Luke Jew,$\!^{1}$\thanks{E-mail: \url{luke.jew@physics.ox.ac.uk}}
Angela C. Taylor$^{1}$
Michael E. Jones,$\!^{1}$
A. Barr,$\!^{2}$
H.\,C. Chiang,$\!^{3}$
\newauthor
C. Dickinson,$\!^{2,4}$
R.\,D.\,P. Grumitt,$\!^{1}$
S.\,E. Harper,$\!^{2}$
H.\,M. Heilgendorff,$\!^{5}$
\newauthor
J. Hill-Valler,$\!^{1}$
J.\,L. Jonas,$\!^{6,7}$
J.\,P. Leahy,$\!^{2}$
J. Leech,$\!^{1}$
T.\,J. Pearson,$\!^{4}$
\newauthor
M.\,W. Peel,$\!^{8,9}$
A.\,C.\,S. Readhead,$\!^{4}$
J. Sievers,$\!^{3}$
\newauthor
\\
$^{1}$Sub-department of Astrophysics, University of Oxford, Denys Wilkinson Building, Keble Road, Oxford OX1 3RH, UK\\
$^{2}$Jodrell Bank Centre for Astrophysics, Alan Turing building, School of Physics and Astronomy, The University of Manchester,\\
Oxford Road, Manchester, M13 9PL, Manchester, U.K. \\
$^{3}$Department of Physics, McGill University, 3600 Rue University, Montr\'{e}al, QC H3A 2T8, Canada \\
$^{4}$Cahill Centre for Astronomy and Astrophysics, California Institute of Technology, Pasadena, CA 91125, USA \\
$^{5}$Astrophysics \& Cosmology Research Unit, School of Mathematics, Statistics \& Computer Science, University of KwaZulu-Natal, \\
$^{6}$Department of Physics and Electronics, Rhodes University, Grahamstown, 6139, South Africa \\
$^{7}$South African Radio Astronomy Observatory, 2 Fir Road, Observatory, Cape Town, 7925 , South Africa \\
$^{8}$Instituto de Astrof\'{i}sica de Canarias, E-38205 La Laguna, Tenerife, Spain\\
$^{9}$Departamento de Astrof\'{i}sica, Universidad de La Laguna (ULL), E-38206 La Laguna, Tenerife, Spain\\
}

\date{Accepted XXX. Received YYY; in original form ZZZ}

\pubyear{2015}

\usepackage{etoolbox}
\makeatletter
\patchcmd\@combinedblfloats{\box\@outputbox}{\unvbox\@outputbox}{}{\errmessage{\noexpand patch failed}}

\begin{document}
\label{firstpage}
\pagerange{\pageref{firstpage}--\pageref{lastpage}}
\maketitle
\begin{abstract}
The cosmic microwave background $B$-mode signal is potentially weaker than the diffuse Galactic foregrounds over most of the sky at any frequency.
A common method of separating the CMB from these foregrounds is via pixel-based parametric-model fitting.
There are not currently enough all-sky maps to fit anything more than the most simple models of the sky.
By simulating the emission in seven representative pixels, we demonstrate that the inclusion of a 5\,GHz data point allows for more complex models of low-frequency foregrounds to be fitted than at present. It is shown that the inclusion of the CBASS data will significantly reduce the uncertainties in a number of key parameters in the modelling of both the galactic foregrounds and the CMB.
The extra data allow estimates of the synchrotron spectral index to be constrained much more strongly than is presently possible, with corresponding improvements in the accuracy of the recovery of the CMB amplitude. However, we show that to place good limits on models of the synchrotron spectral curvature will require additional low-frequency data.
\end{abstract}



\section{Introduction}

The C-Band All-Sky Survey (C-BASS) is a project to produce a high sensitivity all-sky map at 5\,GHz in total intensity and polarization with a resolution of just under $1^\circ$ \citep{Jones2018}. The primary science goal of C-BASS is to be used in combination with other data sets to produce maps of the cosmic microwave background (CMB) that are free from contaminating foreground Galactic emission in both total intensity and polarization. A secondary goal is to make improved measurements of the contaminating components themselves, and in particular to study the structure of the Galactic magnetic field. In this work we test the impact that C-BASS data will have on the measurements of the CMB and foregrounds by fitting parametric models of the sky to simulated data both with and without the C-BASS data point. In addition to C-BASS data we use existing data sets for CMB intensity, and surveys expected in the near future for CMB polarization. 

Although current measurements of the CMB intensity have high sensitivity over a wide range of frequencies and angular scales, there are still degeneracies between  foreground components \citep{PlanckCollaboration2015,Ade2016}. This is due in part to the lack of data at lower frequencies where synchrotron radiation, free-free emission and anomalous microwave emission (AME) can all be significant. Analyses such as \citet{PlanckCollaboration2015} are forced to assume a particular spectral form for the synchrotron emission, and cannot fully discriminate between these three emission mechanisms. The 408 MHz map of \citet{Haslam1982} as reprocessed by \citet{Remazeilles2014} is often used to provide a synchrotron template in total intensity, but has well-known problems with calibration, offsets and image fidelity. The C-BASS intensity survey is designed to provide high-fidelity and well-calibrated maps at a much closer frequency to the other CMB surveys than the 408~MHz map, but with negligible contribution from AME, and thus a significantly different mix of foregrounds to the lower frequency channels of the space-based data sets.  

In polarization, the foregrounds are much simpler, being dominated by synchrotron radiation and dust. However the primordial $B$-mode signal which is the goal of many current observations is relatively much fainter compared to the foregrounds than the intensity or $E$-mode signals, and observations are currently limited by both sensitivity and frequency coverage.
The amplitude of the primordial $B$-mode signal is characterized by $r$, the ratio of amplitudes of tensor to scalar modes in the primordial fluctuation spectrum.
Current limits on $r$ are $r < 0.07$ \citep{BICEP2Collaboration2015}, and the most plausible inflation theories predict that the value of $r$ may be one order of magnitude below this.
At the lower (but plausible) levels of $r$, the $B$-mode signal will be fainter than polarized foregrounds at all frequencies over most of the sky \citep{Dunkley2009}.
It is therefore essential to accurately characterize the polarized foreground emission from our own Galaxy.
In particular, the frequency spectrum of the CMB can be almost degenerate with that of synchrotron radiation at frequencies above the turnover of the CMB spectrum at 217 GHz -- the slope of the CMB spectrum in Rayleigh-Jeans brightness temperature is between $-2$ and $-4$ in the frequency range 200 -- 320 GHz.
In addition, the synchrotron emission could be brighter than the CMB {\em at all frequencies} -- there is no frequency at which synchrotron is negligible if $r \leq 10^{-3}$. 
Accurate estimates of the synchrotron amplitude made at lower frequencies are thus essential to give good subtraction of this foreground \citep{Remazeilles2015}.

In this paper we simulate diffuse Galactic emission in seven pixels in both total intensity and $B$-mode polarization.
The seven pixels were chosen to be representative of a range of foreground environments.
We fit a sky model back to the simulated data both with and without a simulated 5\,GHz data point
and compare the parameter constraints in both cases, in order to demonstrate the impact of the additional data provided by C-BASS.
We also test the impact of mis-modelling the spectral curvature of the synchrotron component (e.g., fitting for a straight spectral index when the model is generated with curvature).
Focussing on a small number of pixels allows a deeper analysis of the subtleties of parameter estimation in this context and the effects of differing relative levels of the various foregrounds. We leave analysis of the whole sky to future work.

This extends on the work presented in Section~7 of \citet{Jones2018} who showed
the impact of the C-BASS data on a single pixel in total intensity
and another pixel in polarization.
In that work they did not consider modelling errors and only used Jeffreys priors on spectral index parameters.
It also extends on the work in Chapter 2 of \citet{Jew2017} who demonstrated the impact of C-BASS in seven pixels and used weakly informative priors on the spectral parameters.
In this work we consider the same seven pixels as \citet{Jew2017}, introduce a modelling error,
and use the full independence Jeffreys-rule prior on the free parameters.
A similar approach was taken by \citet{Hensley2018}, who simulated the parametric fitting process on a single pixel.
They looked specifically at how fitting different dust models with various levels of modelling error changed the biases of the estimated CMB amplitude in the pixel.

The paper is laid out as follows:
In Section~\ref{sec:Models} we describe the spectral models that we use and the frequencies and sensitivities of the simulated observations,
in Section~\ref{sec:Method} we describe the parametric fitting method that we have used,
in Section~\ref{sec:TemperatureResults} we discuss the results from the total intensity pixels,
in Section~\ref{sec:PolarizationResults} we discuss the results from the $B$-mode pixels,
and in Section~\ref{sec:Conclusions} we summarize the results.

\section{Spectral Models and Simulated Pixels}
\label{sec:Models}

In this section we describe the spectral models that we use to simulate the CMB and foregrounds and the frequency channels and sensitivities of the simulated datasets.
We only consider diffuse Galactic synchrotron, free-free, AME and thermal dust emission as foregrounds to the CMB radiation. We do not include compact components such as radio point sources or the Sunyaev-Zeldovich effect since their contributions are negligible on the angular scales of interest.\footnote{Point sources can be dealt with independently either by using high resolution catalogues or statistically in the angular power spectrum.}
We work in units of Rayleigh-Jeans brightness temperature measured in kelvin unless otherwise specified.

We simulate the total intensity and polarization of the emission in seven pixels chosen to
represent a broad range of environments, and an eighth pixel with no foreground contamination. The total intensity signal is constructed from the sum of the CMB, synchrotron, free-free, AME and thermal dust components.
The polarized signal is constructed using only the sum of the CMB, synchrotron and thermal dust components (i.e., neglecting polarized AME and free-free emission). We assume the polarized emission from the Galactic components to be split equally between $E$- and $B$-modes, i.e. we assume that a typical foreground polarized amplitude in $B$ is the same as the typical amplitude in $Q$ or $U$.
We discuss the validity of this approximation in Section~\ref{sec:simulatedPix}.

We do not add realizations of the noise to the simulated data. Instead we use an analytic (Gaussian) form to calculate the likelihood of each simulated observation. The posteriors that we calculate can thus be interpreted as the distribution from which individual realizations of the noisy data would be drawn. This removes the need to calculate many explicit realizations of the data in order to calculate the uncertainty and the bias on the recovered parameters.

\subsection{Spectral models}
\subsubsection{CMB}
The Rayleigh-Jeans brightness temperature spectrum of the CMB, $s_\textrm{CMB}$, has a blackbody spectrum given by 
\begin{equation}
s_{\textrm{CMB}}(\nu) = A_{\textrm{CMB}} \frac{x^2e^x}{(e^x-1)^2},
\end{equation}
where $A_\mathrm{CMB}$ is the amplitude of the CMB fluctuation in the pixel, $x=\left(h\nu\right)/\left(k_\textrm{B}T_\textrm{CMB}\right)$, $h$ is the Planck constant, $k_\textrm{B}$ is the Botlzmann constant, $\nu$ is the frequency and $T_\textrm{CMB}$ is the mean temperature of the CMB, which we take to be 2.7255 K \citep{Fixsen2009}.

\subsubsection{Synchrotron emission}

Over many decades of frequency (100s of MHz up to 100s of GHz) Galactic synchrotron radiation can be approximated as a power-law with temperature spectral index of $\beta\simeq-2.5\-- -3.0$ \citep{Lawson1987,Reich1988,Platania2003,Davies2006,Gold:2008kp,Guzman2011,PlanckCollaboration2014a,PlanckCollaboration2014b,Ade2016}.

Along any one line of sight there are multiple populations of synchrotron-emitting electrons, with each population potentially emitting with a different spectral index.
The frequency spectrum of such a superposition can be (neatly) parametrised using a moment expansion \citep{Chluba2017}.
However, such an expansion introduces more free parameters into the spectral models than there are observations at frequencies that are dominated by synchrotron emission.
Given the small number of low-frequency surveys currently available, instead of a full moment expansion, we consider the inclusion of a simple curvature term in the synchrotron spectral model.
A curved power law corresponds to the line-of-sight average of a Gaussian distribution of spectral indices 
with variance $C_\textrm{s}$ \citep{Chluba2017} and can be parametrized by
\begin{equation}
s_\textrm{s}(\nu) = A_\textrm{s}\left(\frac{\nu}{\nu_0}\right)^{\beta_\textrm{s}+\frac{1}{2}C_\textrm{s}\ln\left(\nu/\nu_0\right)},
\end{equation}
where $A_\textrm{s}$ is the amplitude at a frequency $\nu_0$, $\beta_\textrm{s}$ is the effective spectral index and $C_\textrm{s}$ is the curvature term.

The degree of polarization in synchrotron radiation depends on the spectral index of the electron energy distribution and for typical values of the electron energy spectral index in the galaxy can be up to $\sim$70 percent in ordered magnetic fields \citep{Rybicki1985}.
The interstellar magnetic field has a significant turbulent component and therefore the polarization fraction of diffuse Galactic synchrotron emission will be lower than this across the sky.
At high Galactic latitudes the synchrotron emission is typically up to $\sim 40$\,per cent polarized
\citep{Ade2016,Vidal2014}.
At lower frequencies, and close to the Galactic plane, the synchrotron emission is less polarized due to Faraday depolarization.
In polarization, synchrotron emission is the dominant foreground to the CMB below frequencies around 100\,GHz and so we include it in both our total intensity and polarized models of the Galaxy.

\subsubsection{Free-free emission}

Free-free (or bremsstrahlung) radiation is produced when free electrons scatter off ions in the warm interstellar medium.
The frequency spectrum of free-free emission can be approximated by the two-parameter model of \citet{Draine2011},
\begin{equation}
s_\textrm{ff}(\nu) = T_\textrm{e}\left(1-\exp^{-\tau(\nu)}\right)
\end{equation}
where
\begin{equation}
\begin{aligned}
\tau(\nu) &= 0.05468 T_\textrm{e}^{-3/2}\nu_9^{-2}EM g_\textrm{ff}(\nu),\\
g_\textrm{ff}(\nu) &= \log\left( \exp\left[ 5.960 - \sqrt{3}/\pi \log\left(\nu_9 T_4^{-3/2}\right)\right]+\mathrm{e}\right),
\end{aligned}
\end{equation}
$EM$ is the effective emission measure, $T_\textrm{e}$ is the physical electron temperature of the free-free emitting cloud, $\nu_9$ is the frequency in GHz and $T_4$ is the electron temperature (measured in kelvin) divided by 10,000.

Because the scattering directions in the particle collisions are random, free-free emission is intrinsically unpolarized.
At high angular resolutions free-free emission can be up to 10\,per cent polarized along the edges of bright \ion{H}{ii} regions due to Thomson scattering \citep{Rybicki1985,Keating1998} but elsewhere the 
upper limits are typically $\lesssim 1\,\textrm{per cent}$ \citep{Macellari2011a}.
We therefore ignore polarized free-free emission in this work.

\subsubsection{AME}

AME is an additional component of diffuse Galactic emission, which can be significant in the range of 10s of GHz.
Currently, the most well developed model of AME is spinning dust \citep{Draine1997}.
However, other components such as magnetic dust may contribute \citep{Draine1998a}.
See, for example, \citet{Dickinson2018a} and the references therein for more details.

In this work we consider only a single component of spinning dust.
We model the frequency spectrum of AME with a {\sc spdust2} spectrum \citep{Ali-Haimoud2008,Silsbee2010} that is allowed to shift in logarithmic frequency-brightness space,
with a Rayleigh-Jeans brightness spectrum given by
\begin{equation}
s_\textrm{sd}(\nu) = A_\textrm{AME}\left( \frac{\nu_0}{\nu}\right)^2\frac{F(\nu \nu_\textrm{p0}/\nu_\textrm{peak})}{F(\nu_0 \nu_\textrm{p0}/\nu_\textrm{peak})},
\end{equation}
where $A_\textrm{AME}$ is the amplitude at frequency $\nu_0$, $\nu_\textrm{peak}$ is the peak frequency, $F$ is the template spectrum and $\nu_{\textrm{p}0}$ is the peak frequency of the template. This follows the same prescription as \cite{Bennett2013,PlanckCollaboration2015}.

Theory suggests that AME should only be very weakly polarized.
\citet{Draine2016} predict a polarization fraction of $10^{-6}$ and 
current measurements place upper limits of $\sim1\%$ on the polarization fraction of diffuse AME.
See the review of theory and observations in \citet{Dickinson2018a}.
In this work we do not include a polarized component of AME.

\subsubsection{Thermal dust emission}

Interstellar dust grains radiate thermally. 
The Rayleigh-Jeans brightness spectrum of clouds of inter-stellar dust can be approximated at frequencies below the peak of the emission at $\sim$3\,THz as a modified blackbody spectrum given by
\begin{equation}
s_\textrm{d}(\nu) = A_\textrm{d} \left( \frac{\nu}{\nu_0} \right)^{\beta_\textrm{d}+1} \frac{\exp(\gamma\nu_0)-1}{\exp(\gamma\nu)-1},
\end{equation}
where $\gamma = h/\left(k_\textrm{B}T_\textrm{d}\right)$, $A_\textrm{d}$ is the amplitude of emission at reference frequency $\nu_0$, $T_\textrm{d}$ is the thermal temperature of the dust grain and $\beta_\textrm{d}$ is the emissivity spectral index.
Although in principle there will be multiple populations of thermally emitting dust grains, in this work we only consider one. Others have considered increasingly complex thermal dust models \citep[e.g.,][]{Hensley2018}.

Dust grains are not spherically symmetric and radiate more efficiently along their longer axis.
The asymmetric dust grains will align with the local magnetic field.
This causes thermal dust emission to be polarized.
Typical polarization fractions range from 0 to more than 20\,per cent with a median value of 8\,per cent \citep{Ade2015}.
The polarization fraction is higher along lines-of-sight with lower column density and is therefore greatest when the total intensity emission is weakest.
At frequencies above 100\,GHz thermal dust is the dominant foreground to the CMB in total intensity and polarization and so we include it in both models.

\subsection{Parameter Values} \label{sec:simulatedPix}

We carry out our analysis on seven individual pixels in total intensity,
and polarization, with parameter values that are chosen to provide a representative sample of a wide range of foreground environments (and one of the pixels having no foreground contamination).
The eight pixels chosen here do not represent all possible levels of foreground contamination, which would require a full-sky simulation, but they are representative of the combinations of different foreground amplitudes found across the sky. They thus demonstrate the possible range of component separation results given the observations we assume.

\begin{figure}
\centering
\includegraphics[width=\columnwidth]{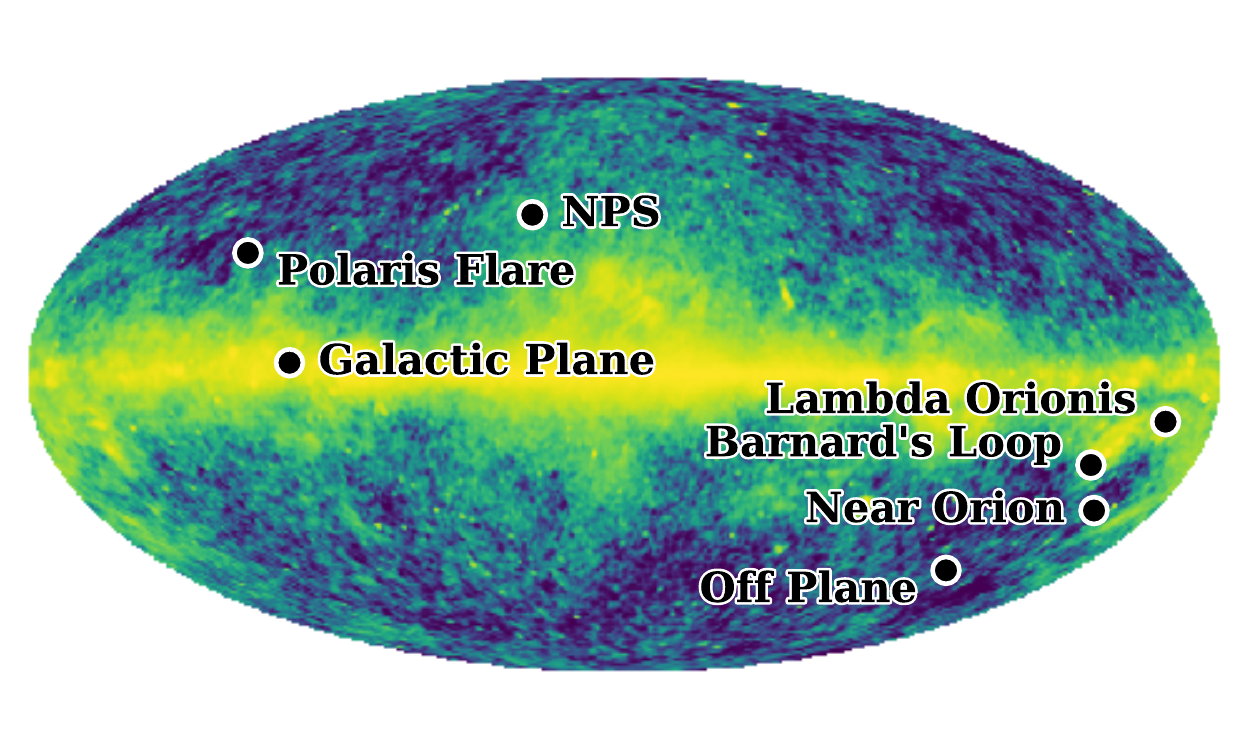}
\caption{\textit{WMAP} \textit{K}-band intensity map with the locations of the pixels considered in this work labelled.}
\label{fig:regionLocations}
\end{figure}

Other than the pixel with no foreground contamination, the foreground amplitude values were selected by picking regions from the \textit{Planck} component maps \citep{PlanckCollaboration2015} and taking the local amplitudes of each component. Other foreground parameters were given the global fiducial values listed below. The locations of the pixels on the sky are shown in Figure~\ref{fig:regionLocations}, with descriptive names corresponding to their positions on the sky. The locations of all the pixels have been observed in the  C-BASS North survey.

\begin{figure}
\centering
\subfloat[Total intensity]{
\label{fig:modelFreqSpectra-I}
\includegraphics[width=\columnwidth]{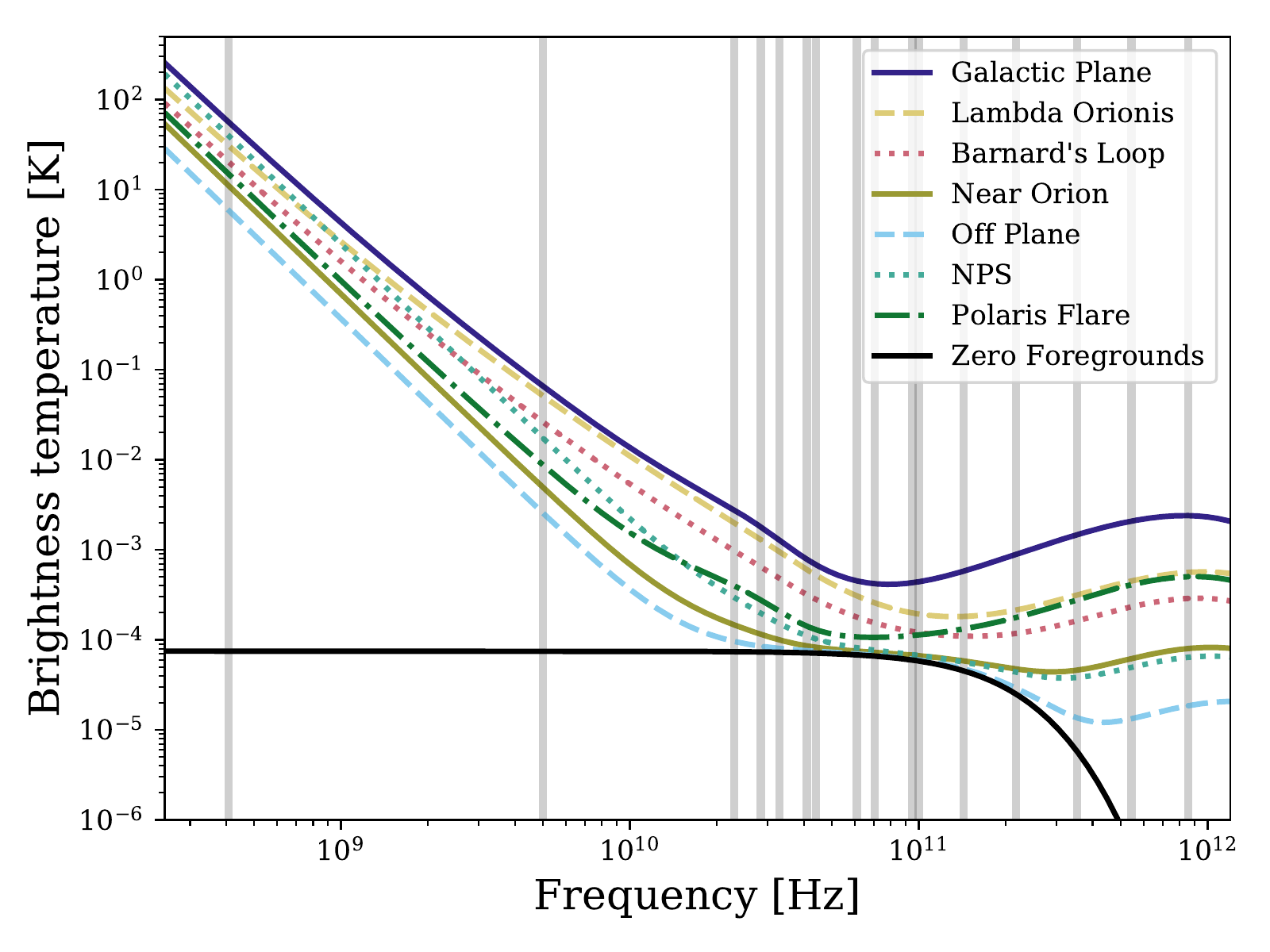}}
\qquad
\subfloat[Polarization]{
\label{fig:modelFreqSpectra-B}
\includegraphics[width=\columnwidth]{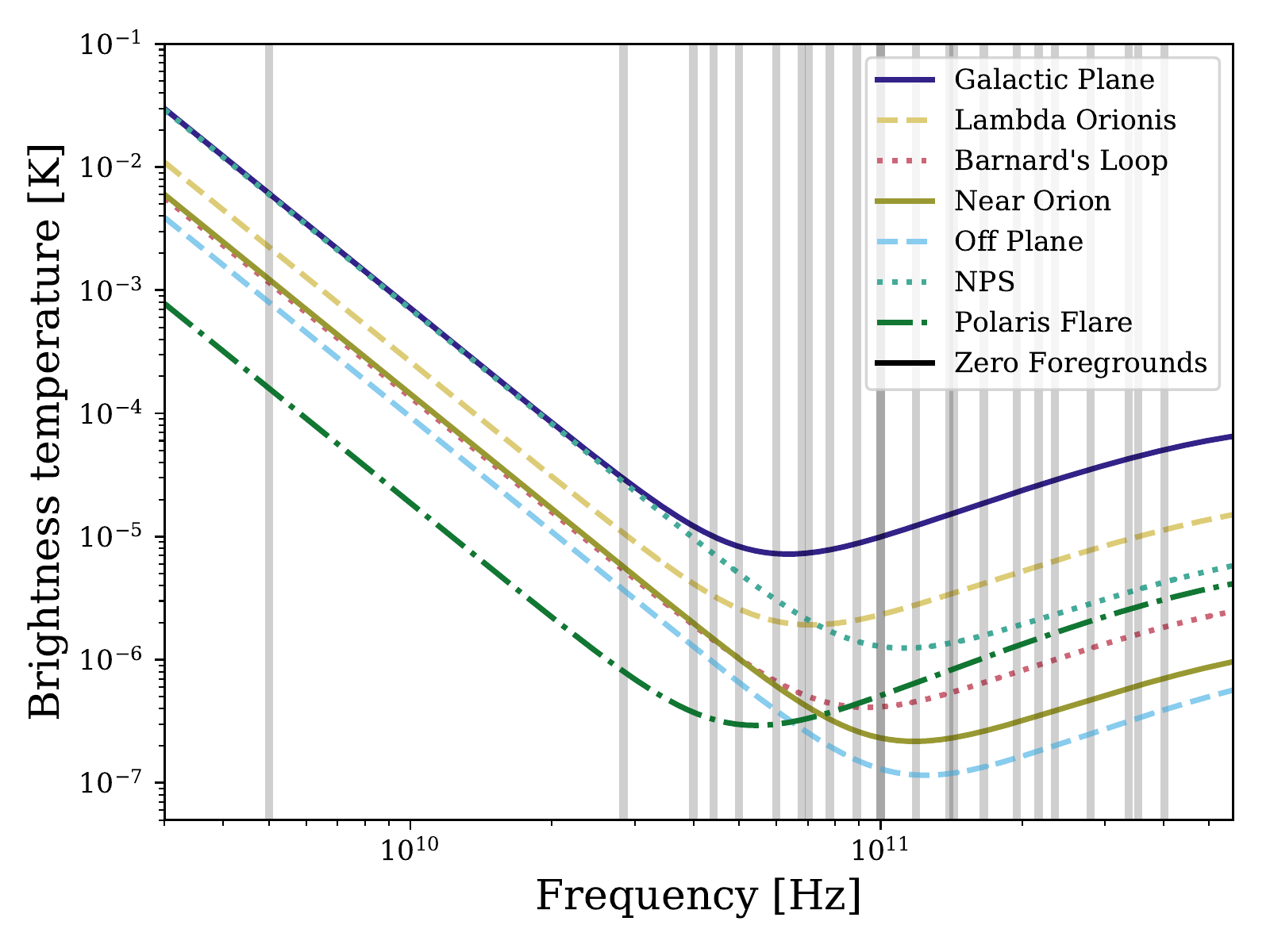}}
\caption{Frequency spectrum of each pixel.
The CMB spectrum is in \textit{solid black} (the $B$-mode signal has been set to zero).
The \textit{vertical grey lines} are at the frequencies of the simulated surveys.}
\label{fig:modelFreqSpectra}
\end{figure}

The parameter values that we use to generate each pixel are listed in Table~\ref{tab:trueParamValsI}, where the superscripts $(I)$ and $(B)$ on the amplitude and reference frequency parameters indicate whether they are for total intensity or polarization pixels.
Specifically;
\begin{itemize}
    \item The synchrotron amplitudes, free-free emission measures and thermal dust parameters
were taken from \citet{PlanckCollaboration2015}.
    \item The polarized amplitudes were set to the estimates of the polarized intensities divided by $\sqrt{2}$.
    \item The synchrotron spectral indices are set to $-3.1$. The synchrotron spectral curvatures are set to 0 (and also 0.15 when explicitly specified in the text). From a spectral index of $-3.1$ at 0.408\,GHz, a spectral curvature of 0.15 results in a spectral index of $-2.68$ at 100\,GHz.
    \item The free-free electron temperature was set to 7000\,K in each pixel.
    \item The AME amplitude was set to the amplitude of the AME-1 component from \citet{PlanckCollaboration2015}
and the peak frequency was set to 25\,GHz.
    \item The CMB amplitude was set to $75\,\mu\textrm{K}$ in total intensity and $0\,\textrm{K}$ in polarization.
\end{itemize}
The total intensity and polarization frequency spectra for each of the pixels are plotted in Figure~\ref{fig:modelFreqSpectra}.

\begin{table*}
\centering
\caption{Parameter values for each pixel. }
\label{tab:trueParamValsI}
\begin{threeparttable}
\begin{tabular}{llrrrrrrrr}
\hline\hline
Component&Parameter &
  \rotatebox{90}{Galactic Plane} & \rotatebox{90}{Lambda Orionis} & \rotatebox{90}{Barnard's Loop} & 
  \rotatebox{90}{Near Orion} & \rotatebox{90}{Off Plane} & \rotatebox{90}{NPS} &
  \rotatebox{90}{Polaris Flare} & \rotatebox{90}{Zero Foregrounds}\\
\hline
\multirow{5}{*}{Synchrotron}
&$A_\textrm{s}^{(I)}$ [K$_\textrm{RJ}$]
  & 47.5 & 22.7 & 16.6 & 11.0 & 5.88 & 39.5 & 14.4 & 0.00\\
&$A_\textrm{s}^{(B)}$ [mK$_\textrm{RJ}$]
  & 6.10& 2.24 & 1.16 & 1.23 & 0.798 & 5.99 & 0.160 & 0.00\\
 &$\beta_\textrm{s}$ & \ldots& \ldots& \ldots&$-3.1$& \ldots&\ldots&\ldots& -\\
 &$C_\textrm{s}$$^\clubsuit$& \ldots & \ldots& \ldots & 0.0 & \ldots&\ldots&\ldots&-\\
 &$\nu_{\textrm{s},0}^{(I)}$$^\Diamond$ [GHz]& \ldots& \ldots& \ldots&\multicolumn{2}{c}{0.408}&\ldots& \ldots& \ldots\\
 &$\nu_{\textrm{s},0}^{(B)}$$^\Diamond$ [GHz]& \ldots& \ldots& \ldots& \multicolumn{2}{c}{ 5 }& \ldots & \ldots& \ldots\\
\\
\multirow{2}{*}{Free-free$^\dag$}
 &$EM$ [cm$^6$pc]
 & 361 & 331 & 152 & 1.59 & $0.00$ & 4.86 & 20.3 & 0.00\\
 &$T_\textrm{e}$$^*$ [K]& \ldots& \ldots& \ldots& 7000& \ldots& \ldots& \ldots & - \\
\\
\multirow{3}{*}{AME$^\dag$}
 &$A_\textrm{AME}$ [$\mu$K$_\textrm{RJ}$]
& 708 & 207 & 85.5 & 22.9 & 0.00 & 49.3 & 167 & 0.00 \\
 &$\nu_\textrm{p}$ [GHz]& \ldots& \ldots& \ldots&25.0&\ldots& \ldots& \ldots & - \\
 &$\nu_{\textrm{AME},0}$$^\Diamond$ [GHz]& \ldots& \ldots& \ldots& \multicolumn{2}{c}{22.8}& \ldots& \ldots& \ldots\\
\\
\multirow{2}{*}{CMB}
 &$A_\textrm{CMB}^{(I)}$ [$\mu$K$_\textrm{RJ}$]& \ldots& \ldots& \ldots& \multicolumn{2}{c}{75}& \ldots& \ldots& \ldots\\
 &$A_\textrm{CMB}^{(B)}$ [$\mu$K$_\textrm{RJ}$]& \ldots& \ldots& \ldots& \multicolumn{2}{c}{0}& \ldots& \ldots& \ldots\\
\\
\multirow{5}{*}{Thermal dust}
 &$A_\textrm{d}^{(I)}$ [$\mu$K$_\textrm{RJ}$]
& 2080 & 448 & 232 & 61.4 & 12.8 & 49.2 & 410 & 0.00\\
 &$A_\textrm{d}^{(B)}$ [$\mu$K$_\textrm{RJ}$]
& 44.8 & 9.98 & 1.61 & 0.614 & 0.335 & 3.72 & 2.70 & 0.00\\
 &$\beta_\textrm{d}$ 
& 1.55 & 1.48 & 1.59 & 1.55 & 1.63 & 1.53 & 1.63 & - \\
 &$T_\textrm{d}$ 
& 17.5 & 21.2 & 19.0 & 21.5 & 24.9 & 21.8 & 18.1 & - \\
 &$\nu_{\textrm{d},0}^{(I)}$$^\Diamond$ [GHz] & \ldots& \ldots& \ldots& \multicolumn{2}{c}{545}& \ldots& \ldots& \ldots\\
 &$\nu_{\textrm{d},0}^{(B)}$$^\Diamond$ [GHz] & \ldots& \ldots& \ldots& \multicolumn{2}{c}{353}& \ldots& \ldots& \ldots\\
\hline
\end{tabular}
\begin{tablenotes}
\item $^*$ Astrophysical fixed parameter, could in principle vary across the sky.
\item $^\Diamond$ Non-astrophysical fixed parameter.
\item $^\dag$ Only in total intensity.
\item $^\clubsuit$ And 0.15 when specified in the text, i.e. when testing the effect of mis-modelling the synchrotron spectrum.
\end{tablenotes}
\end{threeparttable}
\end{table*}

\begin{figure*}
\includegraphics[width=0.8\textwidth]{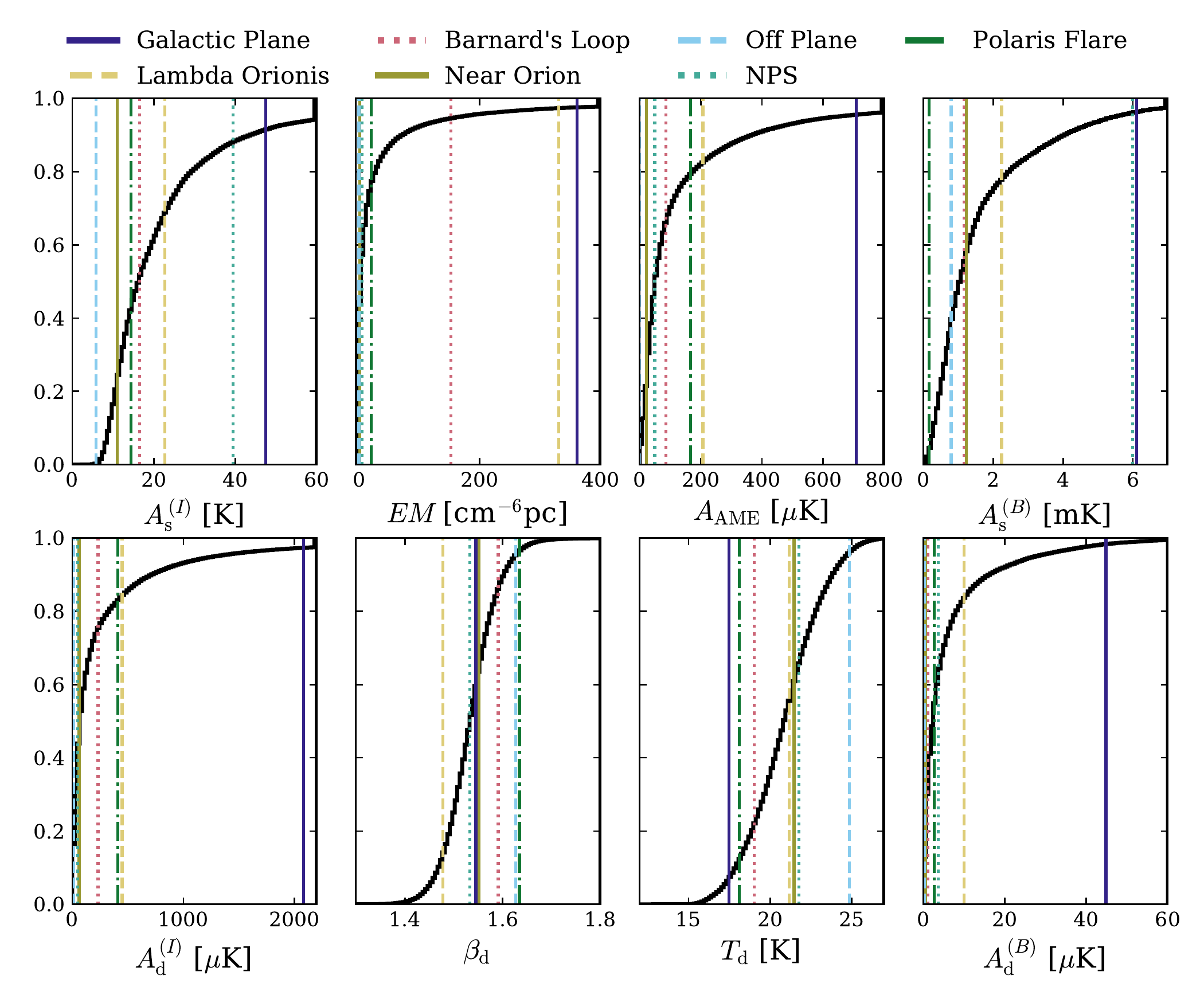}
\caption{Cumulative histograms of the parameter values from the \citet{PlanckCollaboration2015} results (extrapolating the synchrotron amplitude in polarization to 5\,GHz with a temperature spectral index of $-3.1$) with the parameter values of the pixels considered in this work indicated with vertical lines.
}
\label{fig:paramValHist}
\end{figure*}

Figure~\ref{fig:paramValHist} shows cumulative histograms of the parameter values from the Planck 2015 diffuse component separation results \citep{PlanckCollaboration2015},
and the vertical lines are at the parameter values of the pixels listed in Table~\ref{tab:trueParamValsI}.
We extrapolate the polarized synchrotron amplitude from 30\,GHz to our reference frequency of 5\,GHz using a temperature spectral index of $-3.1$ where the spectral curvature is set to zero.

By setting the polarized amplitudes to the polarized intensity divided by $\sqrt{2}$ we have assumed that the polarized synchrotron and thermal dust emission is split equally between the $E$- and $B$-mode components.
Measurements of the $E$ and $B$ spectra of both synchrotron and dust across large areas of the sky suggest that typically the $E$-mode signal is larger by a factor $\sim 2$ \citep{Liu2018,PlanckCollaboration2014}, however this does not qualitatively affect the results presented here.  

We set the CMB polarized amplitude to zero in order to model the situation of attempting to measure a vanishingly small $B$-mode signal with perfect $E$--$B$ separation.
We can then interpret the width of the posterior distribution of the CMB amplitude as the limits on any detection, and any displacement from zero as bias.

\subsection{Frequencies and sensitivities of simulated observations}

We simulate the pixels at frequencies that are characteristic of current and upcoming surveys.
The centre frequencies and sensitivities assigned to each survey are listed in Table~\ref{tab:simDataVals}.
In total intensity the sensitivities correspond to $1^\circ$ pixels.
This results in high signal-to-noise detections of all components (including the CMB) across most of the sky.
In polarization we use sensitivities corresponding to $3^\circ$ pixels.
This scale roughly coincides with the recombination peak and ensures sufficient signal-to-noise to detect the polarized dust emission in all of the \textit{Planck} 353\,GHz pixels.

We have assumed that colour corrections have been made and do not impact the results or errors substantially.
For the total intensity simulations we include the \textit{WMAP}, \textit{Planck} and Haslam all-sky surveys.
This is the same set used in the analysis of \citet{PlanckCollaboration2015}.

For polarization we include the \textit{Planck} surveys and proposed surveys from the next generation space mission \textit{LiteBIRD} \citep{Suzuki2018}.
For \textit{LiteBIRD}, we use the same frequencies and sensitivity values as \citet{Remazeilles2018}.
The sensitivity to $E$- and $B$-mode polarization in the pixels is assumed to be
the same as the sensitivity to Stokes $Q$ and $U$.
These sensitivities are representative of other proposed missions aiming to detect $r\lesssim10^{-3}$.

In future work we will consider a more extensive set of low-frequency surveys, such as Rhodes/HartRAO \citep[total intensity only at 2.3\,GHz][]{Jonas1998}, S-PASS \citep[2.3\,GHz][]{Carretti2019} and QUIJOTE \citep[10--40\,GHz][]{Genova-Santos2015a}.

\begin{table}
\caption{
Frequencies and sensitivities of simulated data, in both intensity and polarization.}
\label{tab:simDataVals}
\begin{threeparttable}
\begin{tabular}{lrrr}
\hline\hline
Name & $\nu$ [GHz] & $\sigma_I$ [$\mu$K\,deg] & $\sigma_P$ [$\mu$K\,deg]\\
\hline
C-BASS $^a$ & 5.0 & 73.0 & 73.0\\
Haslam $^b$ & 0.408 & $2.5\times10^6$ & - \\
\textit{WMAP} K $^c$ & 23 & 5.82 & -\\
\textit{WMAP} Ka & 33 & 4.18 & -\\
\textit{WMAP} Q & 41 & 3.52 & - \\
\textit{WMAP} V & 61 & 3.79 & - \\
\textit{WMAP} W & 95 & 3.92 & -\\
\textit{Planc}k 30 $^d$ & 28.4 & 2.45 & 3.30\\
\textit{Planck} 44 & 44.1 & 2.57 & 3.9 \\
\textit{Planck} 70 & 70.4 & 3.08 & 4.5 \\
\textit{Planck} 100 & 100 & 1.00 & 1.53 \\
\textit{Planck} 143 & 143 & 0.333 & 0.72 \\
\textit{Planck} 217 & 217 & 0.261 & 0.60 \\
\textit{Planck} 353 & 353 & 0.198 & 0.57 \\
\textit{Planck} 545 & 545 & 0.0855 & - \\
\textit{Planck} 857 & 857 & 0.0319 & - \\
\textit{LiteBIRD} 40 $^e$ & 40 & - & 0.613\\
\textit{LiteBIRD} 50 & 50 & - & 0.393\\
\textit{LiteBIRD} 60 & 60 & - & 0.325\\
\textit{LiteBIRD} 68 & 68 & - & 0.265\\
\textit{LiteBIRD} 78 & 78 & - & 0.222\\
\textit{LiteBIRD} 89 & 89 & - & 0.192\\
\textit{LiteBIRD} 100 & 100 & - & 0.150\\
\textit{LiteBIRD} 119 & 119 & - & 0.125\\
\textit{LiteBIRD} 140 & 140 & - & 0.0967\\
\textit{LiteBIRD} 166 & 166 & - & 0.105\\
\textit{LiteBIRD} 195 & 195 & - & 0.0950\\
\textit{LiteBIRD} 235 & 235 & - & 0.125\\
\textit{LiteBIRD} 280 & 280 & - & 0.217\\
\textit{LiteBIRD} 337 & 337 & - & 0.318\\
\textit{LiteBIRD} 402 & 402 & - & 0.615\\
\hline
\end{tabular}
\begin{tablenotes}
\item $^a$ \citet{Jones2018}
\item $^b$ 10\% of median ant temp.
\item $^c$ \citet{Bennett2013}
\item $^d$ \citet{PlanckCollaboration2015}
\item $^e$ Sensitivities from Table 2 of \citet{Remazeilles2018}.
\end{tablenotes}
\end{threeparttable}
\end{table}

\section{Method}
\label{sec:Method}

In this section we describe the fitting algorithm that we use to estimate the parameter posterior distributions, the priors that we have used, and the summary statistics that we calculate from the posterior distributions.

\subsection{Parametric fitting}

We use a Markov-Chain Monte-Carlo (MCMC) method to maximise the posterior distribution of the parameters,
\begin{equation}
p(\vec{\theta}\vert \vec{d}) \propto L(\vec{d}\vert\vec{\theta}) \pi(\vec{\theta})
\end{equation}
where $p(\vec{\theta}\vert \vec{d})$ is the posterior distribution, $L(\vec{d}\vert\vec{\theta})$ is the likelihood, $\pi(\vec{\theta})$ is the prior, $\vec{\theta}$ are the free parameters of the model and $\vec{d}$ are the data \citep{Bayes1763,LaPlace1814}.

To construct our likelihood function we assume that the measurement of a total brightness temperature at each frequency has normally distributed errors about the true temperature, with a variance given by the square of the RMS sensitivity assumed for each measurement. The total likelihood is simply the product of the individual likelihoods across all the frequencies.

To construct the posterior distribution we also need to choose appropriate priors. We want to demonstrate the impact of using different sets of data on the parameter constraints, and therefore we wish to avoid the use of informative priors, which place constraints of the parameter values based on additional information. We note that informative priors are sometimes used to ensure convergence in cases where the data themselves are insufficiently constraining. While this is sometimes a valid choice, here we explicitly want to expose how well the parameters can or can not be constrained by the data.

Flat priors are not always uninformative -- a flat prior in some paramaterizations can induce biases in the posterior distribution.  In single-parameter models, the correct uninformative prior is the Jeffreys prior, which is invariant under a re-parametrization of the likelihood.
The straightforward extension to the multi-parameter case is the multivariate Jeffreys prior, which is the square root of the determinant of the Fisher information matrix $\mathbfss{I}$,
\begin{equation}
\pi_\textrm{MJ}\left( \vec{\theta}\right) = \sqrt{\det \mathbfss{I}(\theta)},
\end{equation}
where the Fisher information matrix is given by
\begin{equation}
\mathbfss{I}(\theta)_{i,j} = -\mathrm{E}\left[
\frac{\partial \log L}{\partial \theta_i} 
\frac{\partial \log L}{\partial \theta_j}
\right],
\end{equation}
and $\mathrm{E}[x]$ is the expectation value of $x$ \citep{Jeffreys1939}.
However, there are well-known problems with the multivariate Jeffreys prior.
For example, when using this prior the maximum posterior estimates of the mean and standard deviation of data that are drawn from a normal distribution
have incorrect degrees of freedom, ($\pi_\textrm{Multivariate Jeffreys}(\mu,\sigma)\propto 1/\sigma^2$). In other cases the multivariate Jeffreys prior introduces significant biases into maximum posterior parameter estimates.
Jeffreys himself advised against its use, and instead suggested the Jeffreys independence rule prior, where each parameter is considered independently in turn \citep{Jeffreys1946}.

For each parameter $\theta_i$, the independence-rule prior is given simply given by
\begin{equation}
\pi(\theta_i) \propto \sqrt{-\mathrm{E}\left[
\left(\frac{\partial \log L}{\partial \theta_i}\right)^2\right] },
\end{equation}
and for the full set of parameters the prior is 
\begin{equation}
\pi_\textrm{JR}(\vec{\theta}) \propto \prod_i \pi(\theta_i).
\end{equation}
The independence-rule Jeffreys prior for each parameter can be derived analytically, and they are listed in Table~\ref{tab:jeffreysPriors}.

Our curved synchrotron spectrum model is not physical for all parameter values at all frequencies. For example, a positively curved power law with falling spectrum will eventually reach a minimum brightness before turning over and rising with frequency.
We therefore impose a joint constraint on the synchrotron spectral index and spectral curvature so that the effective spectral index at 500\,GHz is between $-4$ and $-2$,
\begin{equation}
    -4\leq\beta_\mathrm{s}+\frac{1}{2}C_\mathrm{s}\log\left(\frac{500\,\mathrm{GHz}}{\nu_\mathrm{s,0}}\right)\leq-2.
\end{equation}

\begin{figure*}
    \centering
    \subfloat[Total intensity]{
    \includegraphics[width=0.9\textwidth]{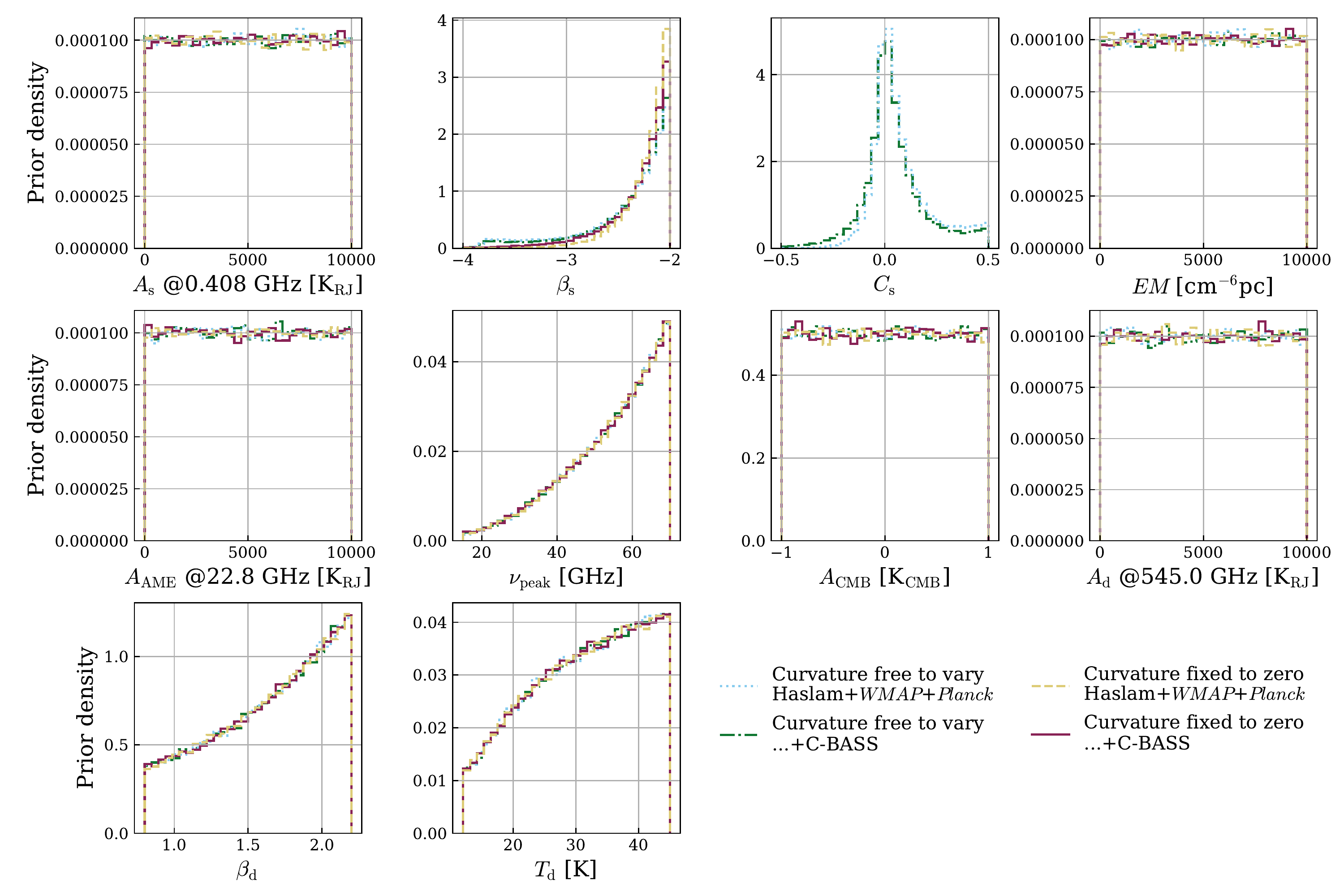}}
    \qquad
    \subfloat[Polarization]{
    \includegraphics[width=0.9\textwidth]{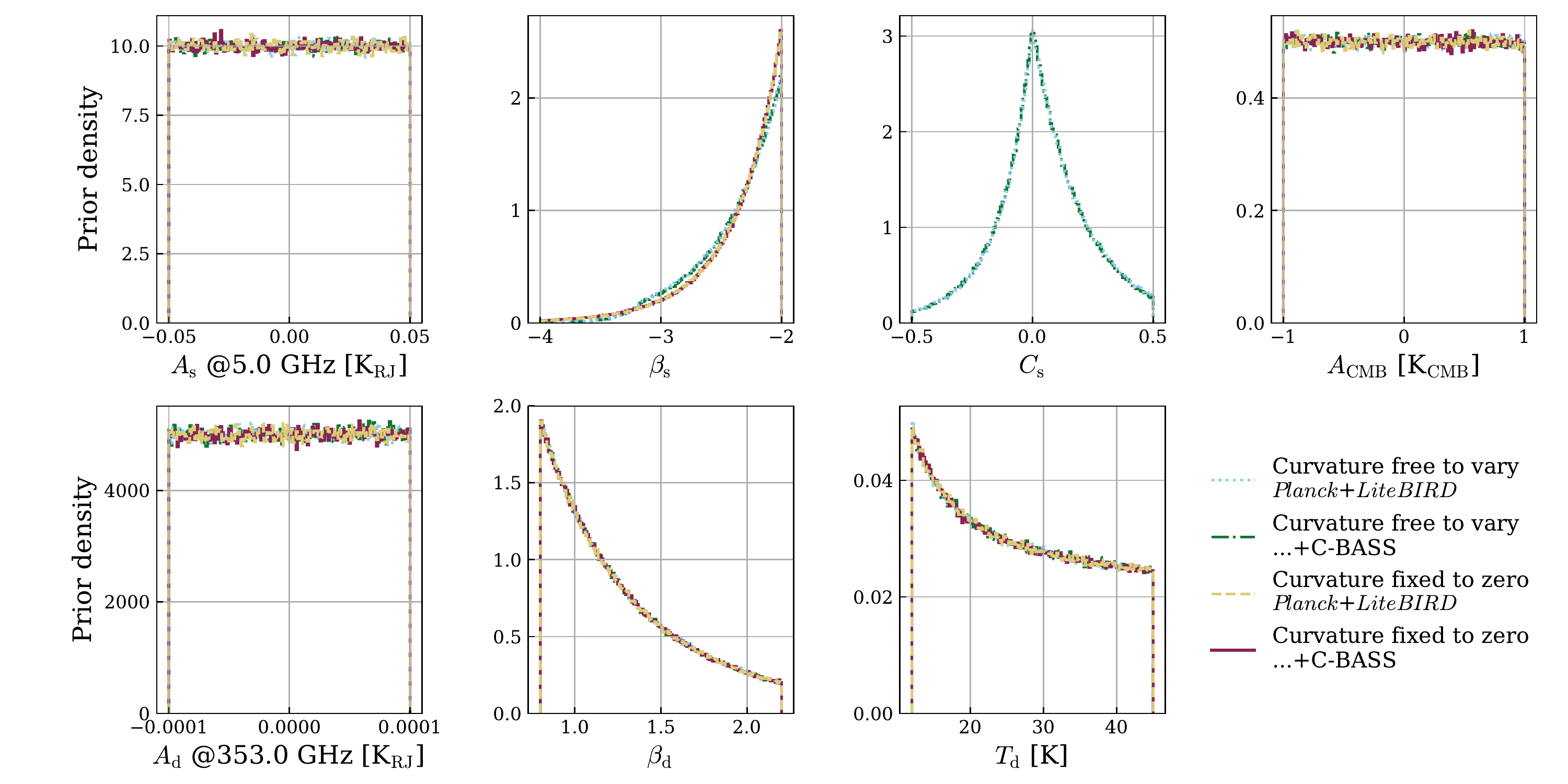}}
    \caption{Marginalized prior distributions of the free parameters in the total-intensity model (\textit{top}) and the polarization model (\textit{bottom}).
    The prior does not change significantly with the combinations of surveys considered in this work.}
    \label{fig:intPriorDistribution}
\end{figure*}

The marginalized prior distributions for each parameter, with and without the C-BASS data point and with and without letting the spectral curvature vary, are shown in Figure~\ref{fig:intPriorDistribution}.
The priors are generally broad and, within the parameter limits, favour values where small changes have the largest effect on the Likelihood.
The synchrotron spectral curvature prior peaks at $C_\textrm{s}=0$, a result of the joint constraint on the spectral index and curvature.
Without the joint constraint, $\pi(C_\textrm{s})$ would increase rapidly with $C_\textrm{s}$.

\begin{table*}
\centering
\caption{Priors on the free parameters. $s_{\textrm{X},i}$ is the brightness temperature of component X in map $i$.
There are two sets of limits listed for the synchrotron and thermal dust amplitude parameters,
the first is for the total intensity case and the second is for the $B$-mode polarization case.
In total intensity we imposed the additional constraint $-4\leq\beta_\mathrm{s}+\frac{1}{2}C_\mathrm{s}\log\left(\frac{500\,\mathrm{GHz}}{\nu_\mathrm{s,0}}\right)\leq-2$.
$F^\prime$ is the derivative (with respect to frequency) of the template spectrum.
The total prior is obtained by multiplying the prior for each parameter together.
}
\label{tab:jeffreysPriors}
\begin{tabular}{lll}
\hline\hline
$\theta$ & $\pi\left(\theta\right)$ & Limits\\
\hline
\multicolumn{3}{c}{\emph{Synchrotron}}\\
$A_\textrm{s}$ & $\propto\,\mathrm{constant}$& $[0,10^4]$\,$\textrm{K}_\textrm{RJ}$, $[-50,50]$\,$\textrm{mK}_\textrm{RJ}$ \\
$\beta_\textrm{s}$
& $\propto\sqrt{ \sum_i\left(\frac{1}{\sigma_i}\frac{s_{\textrm{s},i}}{A_\textrm{s}}\log(\frac{\nu_i}{\nu_0})\right)^2}$
&$[-4,-2]$ \\
$C_\textrm{s}$ 
&$\propto\sqrt{ \sum_i\left(\frac{1}{\sigma_i}\frac{s_{\textrm{s},i}}{A_\textrm{s}}\log^2(\frac{\nu_i}{\nu_0})\right)^2}$
&$[-0.5,0.5]$\\
\\
\multicolumn{3}{c}{\emph{Thermal dust}}\\
$A_\textrm{d}$ & $\propto\,\mathrm{constant}$ & $[0,10^4]$\,$\textrm{K}_\textrm{RJ}$, $[-100,100]$\,$\mu\textrm{K}_\textrm{RJ}$ \\
$\beta_\textrm{d}$ 
&$\propto\sqrt{ \sum_i\left(\frac{1}{\sigma_i}\frac{s_{\textrm{d},i}}{A_\textrm{d}}\log(\frac{\nu_i}{\nu_0})\right)^2}$
&$[0.8,2.2]$ \\
$T_\textrm{d}$ 
&$\propto\sqrt{
\sum_i\left(
\frac{1}{\sigma_i}\frac{s_{\textrm{d},i}}{A_\textrm{d}}
\left[\frac{\nu_0}{1-\exp(-\frac{h\nu_0}{kT_\textrm{d}})}-\frac{\nu_i}{1-\exp(-\frac{h\nu_i}{kT_\textrm{d}})}\right]
\frac{1}{T_\textrm{d}^2}
\right)^2
}$
&$[12,45]$\,K\\
\\
\multicolumn{3}{c}{\emph{Free-free}}\\
$EM$ 
&$\propto\sqrt{\sum_i\left(\frac{1}{\sigma_i}\frac{T_\textrm{e}\tau}{EM}\exp(-\tau)\right)^2}$
&$[0,10^4]$\,cm$^{-6}$pc\\
 & (Note, $\tau\equiv f(T_\textrm{e})\times EM$)\\
 \\
\multicolumn{3}{c}{\emph{Spinning dust}}\\
$A_\textrm{sd}$
&$\propto\,\mathrm{constant}$
&$[0,10^4]$\,$\textrm{K}_\textrm{RJ}$\\
$\nu_\textrm{p}$ 
&$\propto\sqrt{\sum_i\left(\frac{1}{\sigma_i}\frac{s_{\textrm{sd},i}}{A_\textrm{sd}}\frac{\nu_{\textrm{p}0}}{\nu_\textrm{p}^2}\left[\frac{F'(\nu_0\nu_{\textrm{p}0}/\nu_\textrm{p})}{F(\nu_0\nu_{\textrm{p}0}/\nu_\textrm{p})}\nu_0 -  \frac{F'(\nu_i\nu_{\textrm{p}0}/\nu_\textrm{p})}{F(\nu_i\nu_{\textrm{p}0}/\nu_\textrm{p})}\nu_i\right]\right)^2}$
&$[15,70]$\,GHz\\
\\
\multicolumn{3}{c}{\emph{CMB}} \\
$A_\textrm{CMB}$ 
&$\propto\,\mathrm{constant}$
&$[-1,1]$\,$\textrm{K}_\textrm{CMB}$\\
\hline
\end{tabular}
\end{table*}

We are deliberately exploring regimes where it is difficult to constrain all of the parameters with the limited data, and so the choice of prior is important.
An alternative choice of uninformative prior may be the Reference prior, which maximizes the relative entropy between the posterior distribution and the prior \citep{Bernardo1979}. This allows the data to have maximal impact on the posterior.
The Reference priors for highly-dimensional models such as ours are non-trivial to calculate and must be estimated numerically, and we leave it for future work to estimate the reference priors for these models and test whether this provides improved estimates of the parameters.

In both intensity and polarization we both set the synchrotron curvature to zero,
and allow it to vary.
This means that on the simulated data with true synchrotron curvature of 0.15,
we are mis-modelling the synchrotron spectrum when the curvature parameter is set to zero.
We do this to illustrate the effect of using too simple a model that ignores important aspects of the true sky emission.

When fitting the total intensity data we applied a positivity prior on all amplitude parameters.
We relaxed this constraint for the polarization pixels.

We used the Metropolis-Hastings algorithm \citep{Metropolis1953,Hastings1970} to explore the parameter space, implemented in \textsc{PyMC}
\citep{Patil2010}.
The chains were started at the true values for convenience and run for different lengths depending on the number of free parameters.
In polarization we ran the chains for 4 million steps, and during a burn-in period that lasted two hundred thousand steps we tuned the width of the step proposal distribution every one thousand steps, and we thinned the chains by a factor of five.
In total intensity when the curvature was fixed, we ran the chains for ten million steps, and during a burn-in period that lasted for three million steps we tuned the step proposal distribution every one hundred steps, and thinned the resulting chains by a factor of one hundred.
In total intensity when the curvature was free to vary, we ran the chains for one hundred million steps, and had a burn-in period of thirty million steps during which we tuned the step proposal distribution every one hundred steps, and we thinned the resulting chains by a factor of one thousand.
Thinning has no effect on the results it simply reduces the correlation between samples and results in smaller file sizes.

We tested for convergence by inspecting the traces and also using more formal methods.
For each parameter
we used the Raftery-Lewis diagnostic \citep{Raftery1995} to estimate the thinning required to produce an independent chain before
testing for convergence with the Geweke diagnostic test \citep{Geweke1992}.
From preliminary work we found that the total intensity pixels required significantly longer chains than the $B$-mode pixels to strictly pass the convergence tests.
This is because the total intensity pixels have a greater number of correlated and weakly constrained parameters than the $B$-mode pixels.
Shorter chains could be used along with more efficient sampling algorithms such as the No-U-Turn Sampler \citep{Hoffman2011}.
The chains would also converge more quickly if we used informative priors.

\begin{figure*}
    \centering
    \subfloat[Total intensity]{\includegraphics[height=0.27\textheight]{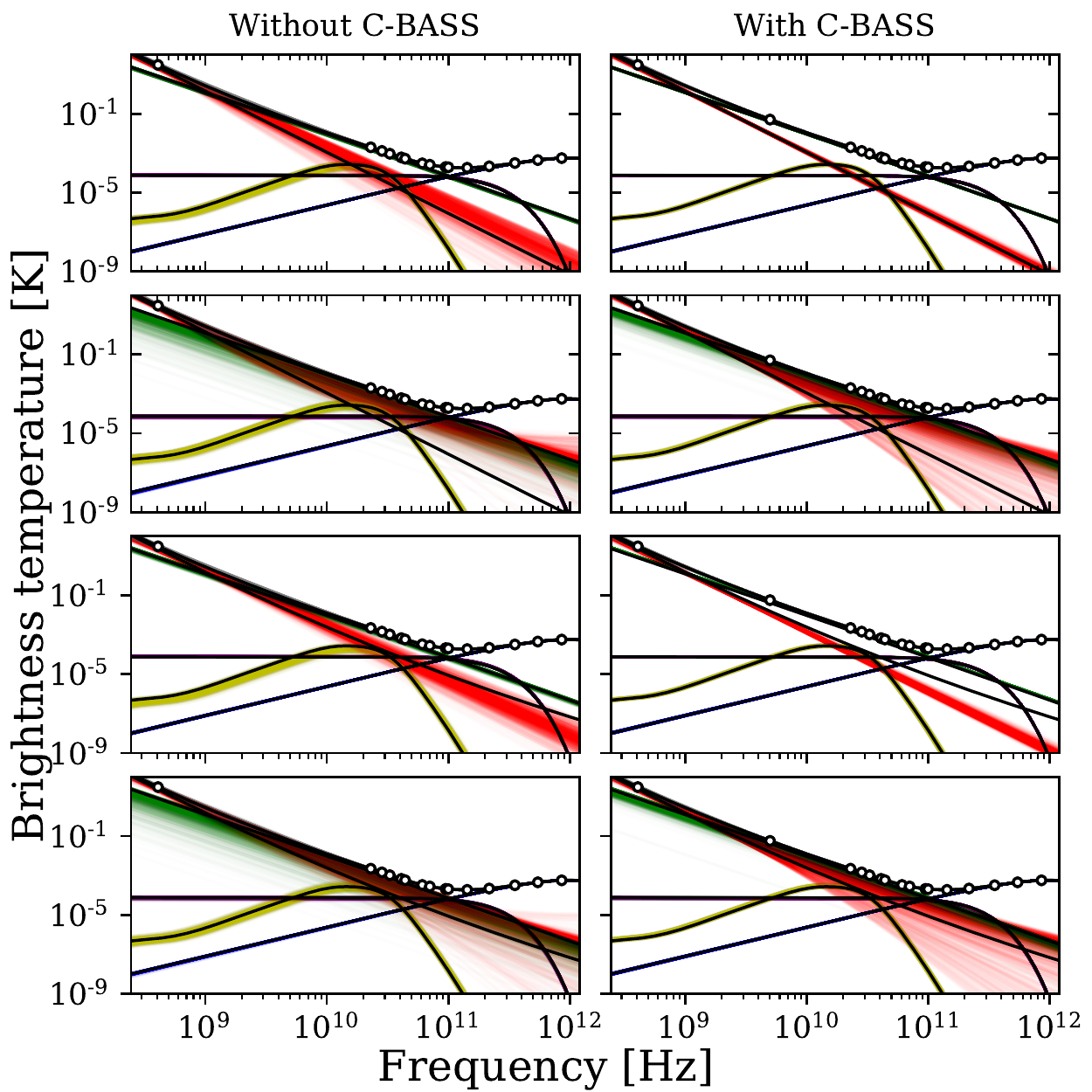}}
    \subfloat[Polarization]{\includegraphics[height=0.27\textheight]{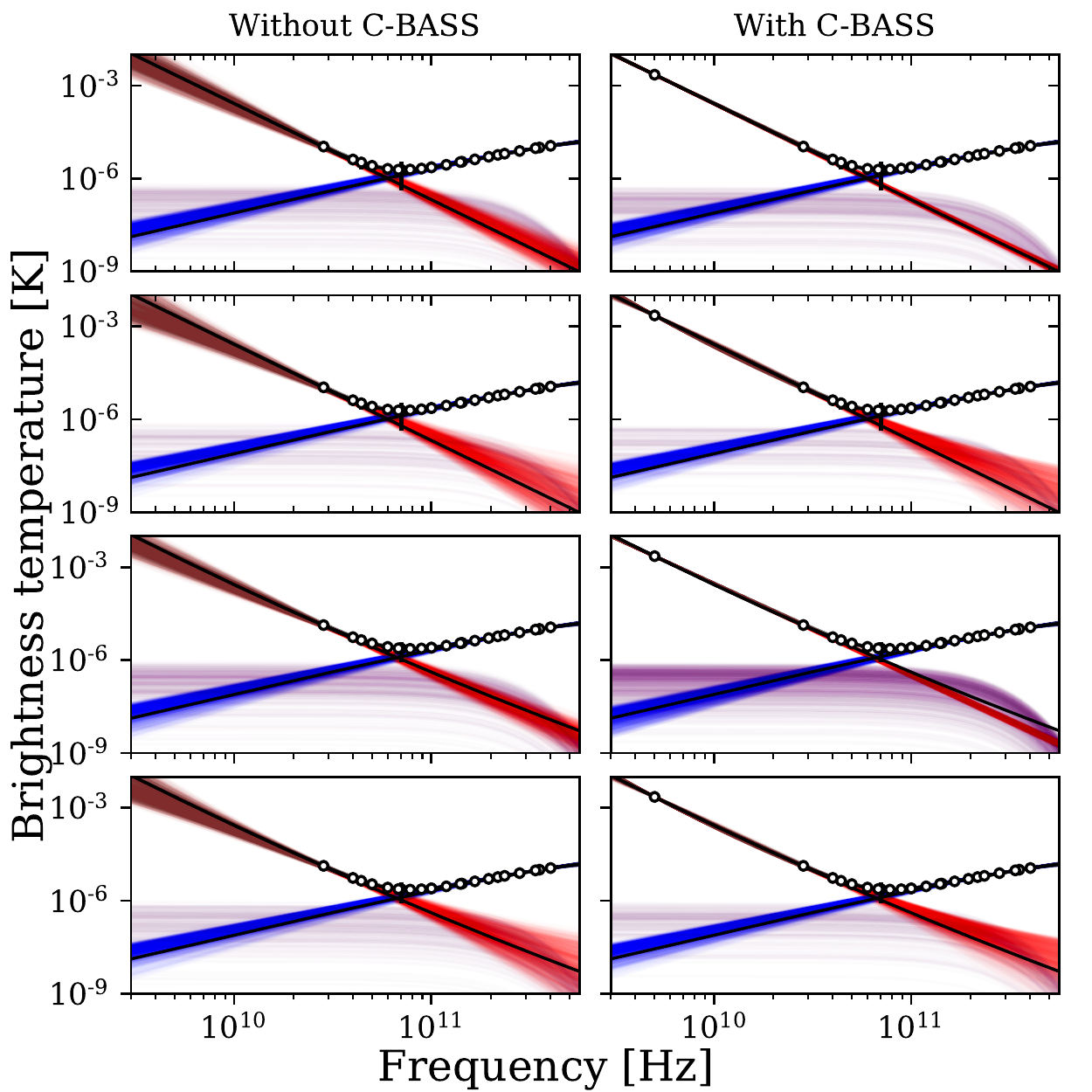}}
    \includegraphics[height=0.27\textheight]{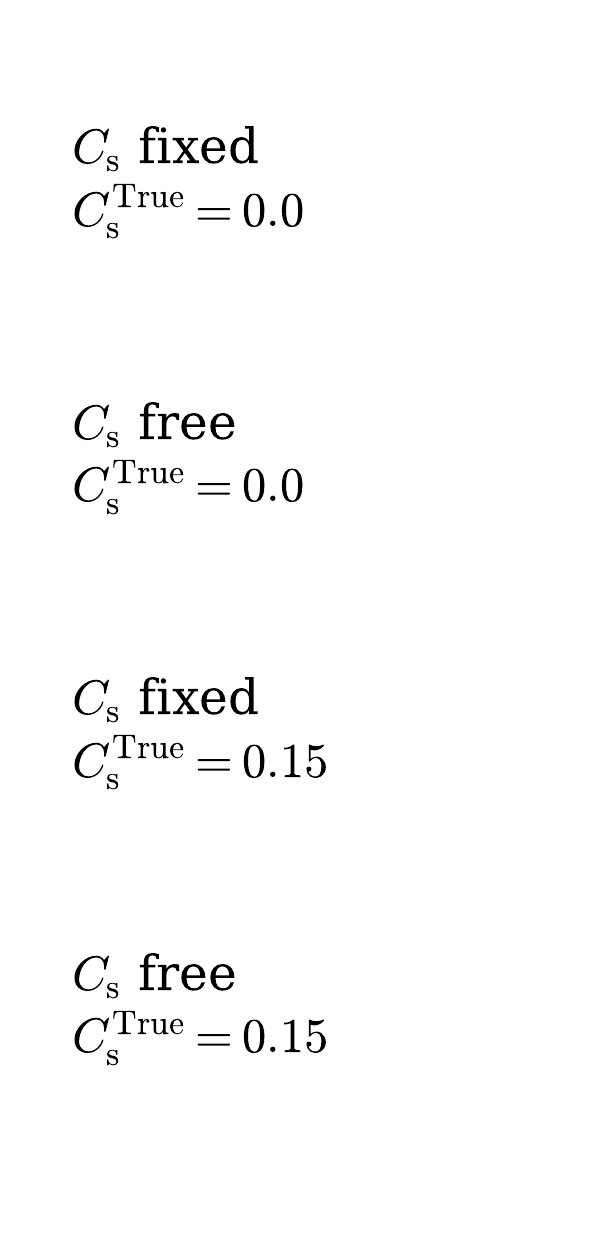}
    \caption{Frequency spectra of a thinned subset of samples from the converged MCMC chains in the Barnard's Loop pixel for both total intensity and polarization with free and fixed synchrotron spectral curvature and true curvatures of 0.0 and 0.15.
The \textit{red lines} are synchrotron, the \textit{blue lines} are thermal dust, the \textit{green lines} are free-free, the \textit{yellow lines} are AME and the \textit{purple lines} are the CMB.
The total signal is shown by the \textit{grey lines}, the true spectra are shown in \textit{black}.}
    \label{fig:freqSpectraExample}
\end{figure*}

As an example, Figure~\ref{fig:freqSpectraExample} shows thinned subsets of the converged chains for the Barnard's Loop pixel in both total intensity and polarization, with free and fixed synchrotron spectral curvature in the fitting, and true curvatures of 0.0 and 0.15.
To condense the complicated multidimensional data to summary statistics
we estimate the covariance of the parameters from their true values.
For the parameters $\theta_i$ and $\theta_j$ the covariance is
\begin{equation}
    \matr{C}_{i,j} = \textrm{E}\left[(\theta_i-\hat{\theta}_i)(\theta_j-\hat{\theta}_j) \right],
\end{equation}
where $\hat{\theta}_i$ is the true value of parameter $\theta_i$.
The total error volume is the determinant of this matrix.
We can compare the error volumes without and with C-BASS by taking their ratios.
Ratios greater than unity indicate an improvement in the total error volume.

The ratios of the total error volumes condense all of the multi-dimensional posterior distributions into a single dimensionless number. 
To investigate the impact on individual parameters we take the square root of the diagonal elements of the covariance matrix.
Assuming that there is no irreducible error, then the total error, $\Delta_i$, on parameter $\theta_i$ is the sum of the bias and variance of the posterior distribution,
\begin{equation}
\Delta_i^2 = \matr{C}_{i,i}^2 = \textrm{E}\left[  (\theta_i-\hat{\theta}_i)^2  \right] = \textrm{Bias}^2\left[\theta\right] + \textrm{Var}\left[\theta\right],
\label{eqn:totErr}
\end{equation}
where the bias and variance functions have their usual definitions;
\begin{align}
    \textrm{Bias}[\theta] &= \textrm{E}[\theta-\hat{\theta}]\\
    \textrm{Var}[\theta] &= \textrm{E}[\theta^2]-\textrm{E}[\theta]^2.
\end{align}

In the same way that we take the ratio of the total error volumes to quantify the impact of the C-BASS data point, we take the ratios of the total errors for individual parameters.
Ratios greater than unity indicate that the parameter constraint has been improved.
Ratios less than unity indicate that the parameter constraint has worsened.

\section{Total intensity results}
\label{sec:TemperatureResults}

In this section we discuss the parameter estimates when fitting the parametric model to the
$1^\circ$ total intensity pixels.
The ratios of the error volumes are listed in Table~\ref{tab:ratioCovDet}.
In the following sections the ratios of the total errors on individual parameters are listed in Table~\ref{tab:totErrRatios}.

\begin{table}
    \centering
    \caption{Ratios of the total error volumes for the total intensity pixels (\textit{top}) and polarization pixels (\textit{bottom}).
    Ratios greater than unity indicate a reduction in the total error volume by the inclusion of the C-BASS data.
The total error volumes were calculated from two sets of simulated data,
with the true synchrotron curvature set to either 0 or 0.15.
In the fitting process the synchrotron curvature parameter was either fixed to zero or allowed to vary freely.
This introduces a modelling error in the case of simulated data with true curvature of 0.15 and when
fixing the curvature to zero in the fitting.}
    \label{tab:ratioCovDet}
    \begin{tabular}{lrrrr}
    \hline\hline
    True $C_\mathrm{s}$ value & \multicolumn{2}{c}{0.0} & \multicolumn{2}{c}{0.15} \\
    $C_\mathrm{s}$ free or fixed & free & fixed & free & fixed \\
    \hline
    \multicolumn{5}{c}{Total intensity}\\
    \hline
    Galactic Plane & 9,000 & 2,000,000 & 5,000 & 4,000\\
    Lambda Orionis & 1,000,000 & 500,000 & 6,000 & 2,000\\
    Barnard's Loop & 4,000,000 & 600,000 & 300,000 & 3,000\\
    Near Orion     & 300 & 300 & 2,000 & 5\\
    Off Plane      & 1,000 & 7,000 & 4,000 & 200\\
    NPS            & 70,000 & 1,000,000 & 30,000,000 & 20,000\\
    Polaris Flare  & 300,000 & 10,000 & 200,000 & 50\\
    \\
    Geometric mean& 50,000 & 70,000 & 100,000 & 600\\
    \hline
    \multicolumn{5}{c}{Polarization}\\
    \hline
    Galactic Plane & 2,000,000 & 200,000 & 2,000,000 & 60,000\\
    Lambda Orionis & 90,000 & 60,000 & 100,000 & 2,000\\
    Barnard's Loop & 10,000 & 100,000 & 10,000 & 7,000\\
    Near Orion     & 10,000 & 100,000 & 20,000 & 4,000\\
    Off Plane      & 5,000 & 60,000 & 5,000 & 6,000\\
    NPS            & 2,000,000 & 10,000 & 2,000,000 & 600\\
    Polaris Flare  & 50 & 700 & 60 & 1,000\\
    \\
    Geometric Mean & 10,000 & 20,000 & 10,000 & 3,000\\
    \hline

    \end{tabular}
\end{table}

In Section~\ref{sec:tempStraightSynch} we consider the case where 
the synchrotron spectral curvature is fixed to the true value
of zero during the fitting.
The C-BASS data point only has a small impact on the dust parameters and we focus our discussion
on the low-frequency foreground parameters.
In Section~\ref{sec:tempCurvedSynch} we present the results when the curvature is allowed to vary.
In Section~\ref{sec:tempModErr} we introduce a modelling error by setting the true   synchrotron curvature to 0.15 but fix its value to zero in the fitting.

\subsection{Straight synchrotron spectrum}
\label{sec:tempStraightSynch}
\begin{figure}
\centering
\includegraphics[width=\columnwidth]{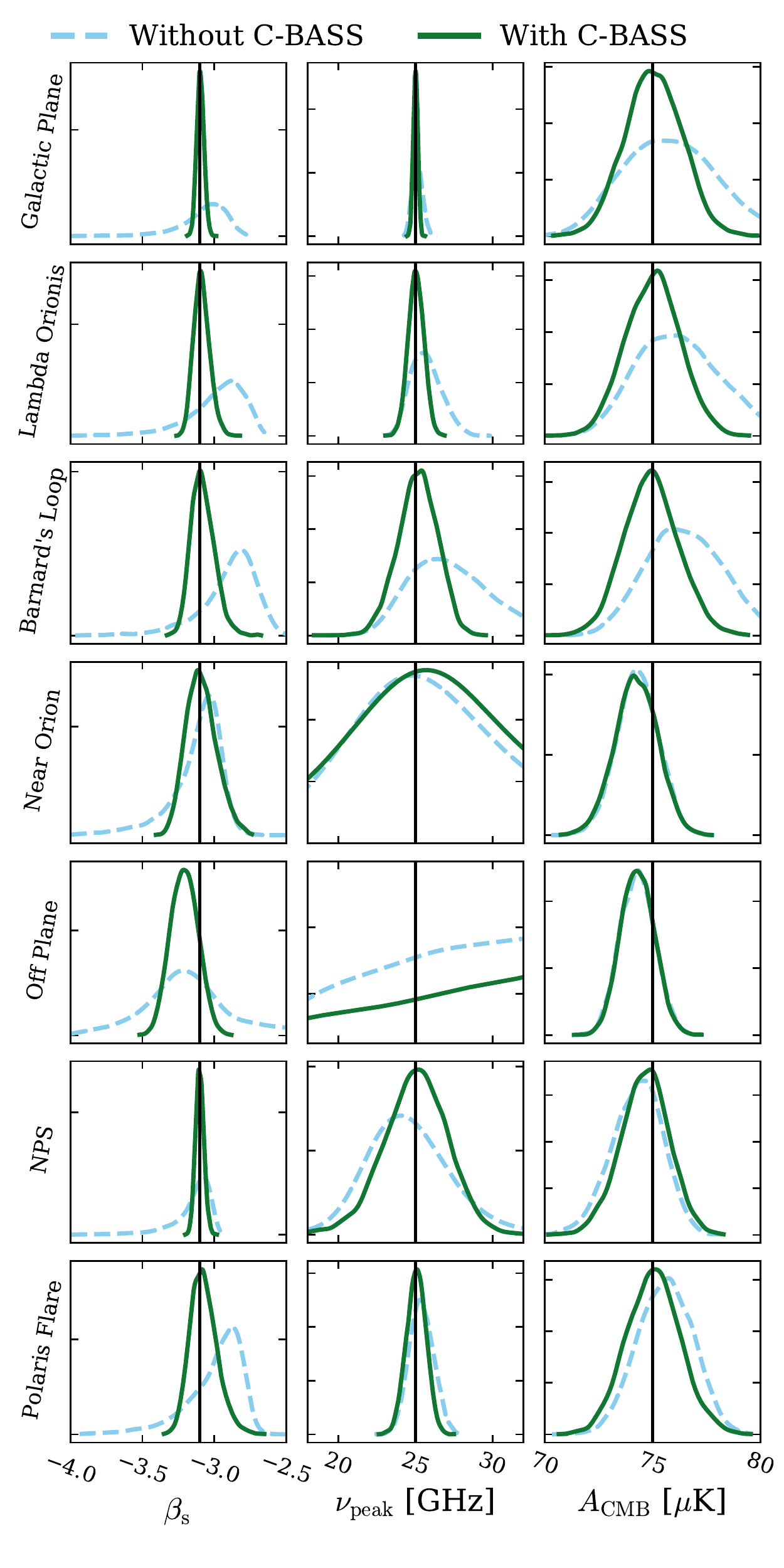}
\caption{Marginalised PDFs of the total intensity low-frequency spectral parameters ($\beta_\mathrm{s}$ and $\nu_\mathrm{peak}$) and the CMB amplitude ($A_\mathrm{CMB}$) that were obtained when fitting the model to the data without C-BASS (\textit{dashed cyan}) and with C-BASS (\textit{solid green}) when the synchrotron spectral curvature was fixed to its true value of zero.
}
\label{fig:pdf_I_lowFreq_straight}
\end{figure}

First we consider the results when the curvature parameter is fixed to its true value of zero.
The marginalised PDFs of the low-frequency spectral parameters (synchrotron spectral index and AME peak frequency) and the CMB amplitude are shown in Figure~\ref{fig:pdf_I_lowFreq_straight}.

Without the C-BASS data point (\textit{dashed cyan} lines) the synchrotron spectral index models cannot be convincingly constrained.
The shallowest spectral indices are excluded by the data and the lower bound on its steepness is set by the prior distribution.
In most of the pixels the synchrotron spectral index posterior distribution does not peak at the true value of $-3.1$.
When the C-BASS data point is included (\textit{solid green} lines), the synchrotron spectral index is well constrained in all pixels, although there remains a bias at the $1\sigma$ level in the lowest-foreground Off Plane pixel.

In all pixels with non-zero AME amplitude (all except the Off Plane pixel), the constraint on the AME peak frequency is also improved by the inclusion of the C-BASS data point.

Because of degeneracies between parameters, without the C-BASS data point the estimates of the CMB amplitude are slightly biased at the sub $1\sigma$ level in many of the pixels and including C-BASS reduces these biases in the pixels with brightest foreground emission.

\begin{figure*}
\centering
\includegraphics[width=0.2\columnwidth]{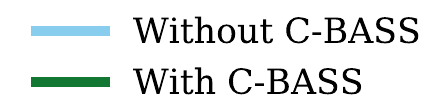}
\includegraphics[width=\textwidth]{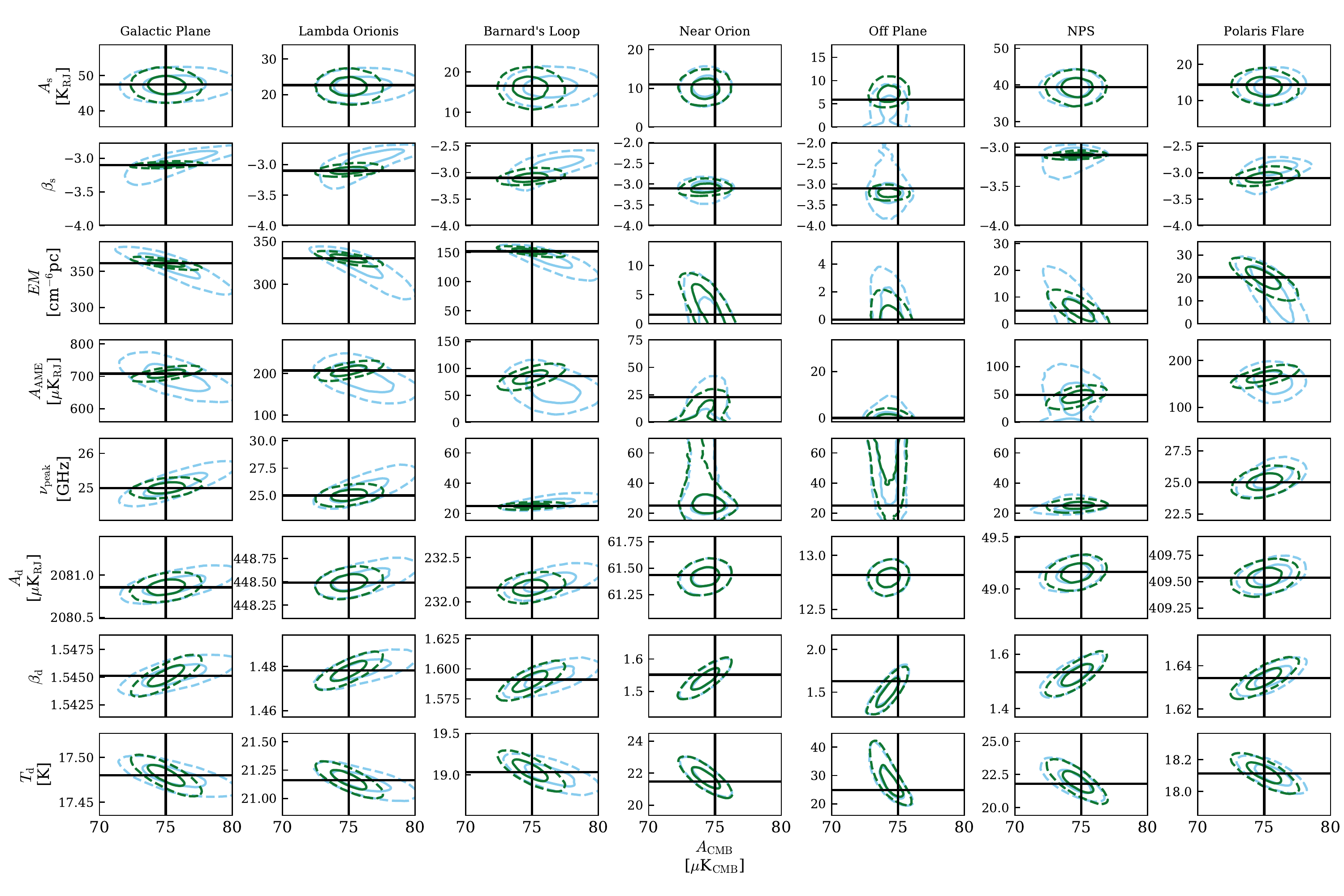}
\caption{The covariance between the CMB amplitude an the other free parameters in the total intensity pixels when the synchrotron spectral curvature is fixed to its true value of zero in the fitting.
The contours show the 1- and 2-$\sigma$ levels in \textit{solid lines} and \textit{dashed lines} respectively.
The \textit{cyan lines} are from the chain without C-BASS.
The \textit{green lines} are from the chain with C-BASS.
The \textit{black lines} are the true parameter values.
}
\label{fig:I_straight_allCmbCov}
\end{figure*}

The covariances between the amplitude of the CMB and other foreground parameters are shown in Figure~\ref{fig:I_straight_allCmbCov}.
Without C-BASS, in the high-foreground pixels (Galactic Plane, Lambda Orionis and Barnard's Loop) degeneracies between the low-frequency components and the CMB slightly bias the amplitude of the CMB high.
The C-BASS data point breaks these degeneracies and removes the bias on the CMB amplitude.

In lower-foreground pixels (Near Orion, Off Plane, NPS and Polaris Flare) the degeneracies between the low-frequency foreground parameters are smaller and so the C-BASS data have a smaller impact of the CMB amplitude.
Degeneracies between the CMB and dust parameters result in small ($<1\sigma$) biases in the estimates of the CMB amplitude to values closer to zero.
The C-BASS data have negligible impact on these parameters in these pixels and so can not remove the bias on the CMB.

We quantify the improvement 
that including the C-BASS data has on the parameter constraints
by taking the ratios of the total errors on the marginalized parameter estimates when not including C-BASS to the total errors when it is included.
These ratios are then the improvement factors, and factors greater than unity indicate an improvement.
The improvement factors are listed in Table~\ref{tab:totErrRatios}.

The total error on the synchrotron amplitude (at 408\,MHz) is not affected by the inclusion of the\
 5\,GHz data
point as this is already constrained by the 408\,MHz Haslam map.
C-BASS only has an impact on this parameter in the Near Orion, Off Plane pixel and Zero Foreground pixels (i.e. those with lowest foreground contamination).

Neglecting the Zero Foregrounds pixel, the improvement factors for the synchrotron spectral index are between 1.8 and 6.4.
Unsurprisingly, C-BASS has the biggest impact when the synchrotron foregrounds is brightest.
The synchrotron amplitude at 408\,MHz
is strongest in the Galactic Plane pixel and then in decreasing order in the NPS, Lambda Orionis, Barnard's Loop,
Polaris Flare, Near Orion, Off Plane and Zero Foregrounds pixels respectively.

C-BASS also allows tighter constraints to be placed on the free-free and AME parameters.
The total error on the free-free emission measure is improved in all pixels (except Near Orion where it has no impact) by the inclusion of C-BASS, with improvement factors between 1.4 and 5.1.
The improvement is greatest in the three pixels with highest foreground contamination (Galactic Plane, Lambda Orionis and Barnard's Loop).

When the AME amplitude is non-negligible, C-BASS improves the total error on the peak frequency.
When the AME amplitude is negligible the total error on the peak frequency increases with the inclusion of C-BASS
because as the amplitude is more tightly constrained to zero,
the peak-frequency parameter
can explore its prior more freely with minimal impact on the posterior.

The AME parameters could likely be much more strongly constrained by fixing the free-free emission measure.
In practice this would be equivalent to assuming the free-free emission could be removed using, for example, an H$\alpha$ template and correctly accounting for any artefacts in those templates \citep{Dickinson2003,Finkbeiner2003}.
Additional low-frequency observations (e.g. at 10--30\,GHz) would also improve the constraints on the low-frequency foreground parameters.

\begin{table*}
\centering
\caption{Improvement factors on the model parameters with no modelling error for total intensity
when the synchrotron curvature is fixed to its true value of zero during fitting (\textit{first section}) and when the curvature is free to vary (\textit{second section}), and in for the polarization when the curvature is fixed to zero during fitting (\textit{third section}) and when it is free to vary (\textit{fourth section}).
The table contains ratios of the total marginalized error on each parameter when not including
a C-BASS data point to the error when including the C-BASS data point.
Ratios greater than unity indicate improvements in the constraints (highlighted in bold face).
}
\label{tab:totErrRatios}
\begin{tabular}{llrrrrrrrr}
\hline\hline
Component & Parameter & 
\rotatebox{70}{Galactic Plane}& \rotatebox{70}{Lambda Orionis} & \rotatebox{70}{Barnard's Loop} &
\rotatebox{70}{Near Orion} & \rotatebox{70}{Off Plane} & \rotatebox{70}{NPS} & 
\rotatebox{70}{Polaris Flare} & \rotatebox{70}{Zero Foregrounds} \\
\hline
\multicolumn{10}{c}{Total intensity, curvature fixed to zero}\\
\hline
\multirow{2}{*}{Synchrotron}
&$A_\textrm{s}$
&1.0
&1.0
&1.0
&\textbf{1.1}
&\textbf{1.4}
&1.0
&1.0
&\textbf{2.3}
\\

&$\beta_\textrm{s}$
&\textbf{6.4}
&\textbf{4.0}
&\textbf{3.3}
&\textbf{1.8}
&\textbf{2.6}
&\textbf{5.4}
&\textbf{2.3}
&-
\\
\\
Free-free
&$EM$
&\textbf{4.4}
&\textbf{4.6}
&\textbf{5.1}
&1.0
&\textbf{1.8}
&\textbf{1.9}
&\textbf{2.1}
&\textbf{1.4}
\\
\\
\multirow{2}{*}{AME}
&$A_\textrm{AME}$
&\textbf{3.3}
&\textbf{2.9}
&\textbf{2.7}
&\textbf{1.1}
&\textbf{2.1}
&\textbf{2.4}
&\textbf{2.0}
&0.9
\\
&$\nu_\textrm{peak}$
&\textbf{2.3}
&\textbf{2.3}
&\textbf{3.6}
&\textbf{1.2}
&0.9
&\textbf{3.1}
&\textbf{1.4}
&-
\\
\\
CMB
&$A_\textrm{CMB}$
&\textbf{1.8}
&\textbf{1.8}
&\textbf{1.9}
&1.0
&1.0
&\textbf{1.2}
&\textbf{1.1}
&\textbf{1.1}
\\
\\
\multirow{3}{*}{Thermal dust}
&$A_\textrm{d}$
&\textbf{1.3}
&\textbf{1.4}
&\textbf{1.4}
&1.0
&1.0
&1.0
&\textbf{1.1}
&1.0
\\
&$\beta_\textrm{d}$
&1.0
&\textbf{1.1}
&\textbf{1.1}
&1.0
&1.0
&1.0
&1.0
&-
\\
&$T_\textrm{d}$
&1.0
&\textbf{1.1}
&\textbf{1.1}
&1.0
&1.0
&1.0
&1.0
&-
\\
\hline
\multicolumn{10}{c}{Total intensity, curvature free to vary}\\
\hline
\multirow{3}{*}{Synchrotron}
&$A_\textrm{s}$
&\textbf{1.4}
&\textbf{1.4}
&1.0
&1.0 
&\textbf{1.8}
&1.0
&1.0
&\textbf{2.4}
\\

&$\beta_\textrm{s}$ 
&\textbf{2.4}
&\textbf{2.3}
&\textbf{2.4}
&\textbf{1.9}
&\textbf{1.1}
&\textbf{2.5}
&\textbf{2.7}
&-
\\

&$C_\textrm{s}$
&\textbf{1.1}
&\textbf{1.3}
&\textbf{1.5}
&\textbf{1.4}
&1.0
&\textbf{1.4}
&\textbf{2.0}
&-
\\
\\
Free-free
&$EM$
&\textbf{1.7}
&\textbf{1.4}
&\textbf{1.1}
&0.7
&\textbf{1.1}
&0.7
&\textbf{1.4}
&\textbf{1.4}
\\
\\
\multirow{2}{*}{AME}
&$A_\textrm{AME}$
&\textbf{1.8}
&\textbf{1.6}
&\textbf{2.3}
&\textbf{1.2}
&\textbf{1.3}
&\textbf{1.3}
&\textbf{2.0}
&0.9
\\

&$\nu_\textrm{peak}$
&\textbf{2.1}
&\textbf{2.1}
&\textbf{2.6}
&\textbf{1.5}
&0.9
&\textbf{3.7}
&\textbf{1.8}
&-
\\
\\
CMB
&$A_\textrm{CMB}$
&\textbf{1.2}
&\textbf{1.2}
&\textbf{1.9}
&1.0
&\textbf{1.5}
&1.0
&\textbf{2.9}
&\textbf{1.1}
\\
\\
\multirow{3}{*}{Thermal dust}
&$A_\textrm{d}$
&\textbf{1.4}
&\textbf{2.3}
&\textbf{4.3}
&\textbf{1.7}
&\textbf{2.2}
&\textbf{1.1}
&\textbf{14.0}
&1.0
\\

&$\beta_\textrm{d}$ 
&\textbf{1.2}
&\textbf{1.5}
&\textbf{2.0}
&1.0
&1.0
&1.0
&\textbf{1.5}
&-
\\

&$T_\textrm{d}$
&\textbf{1.1}
&\textbf{1.4}
&\textbf{1.6}
&1.0
&1.0
&1.0
&\textbf{1.2}
&-
\\
\hline
\multicolumn{10}{c}{Polarization, curvature fixed to zero}\\
\hline
\multirow{3}{*}{Synchrotron}
&$A_\textrm{s}$
&\textbf{62.1} 
&\textbf{48.8} 
&\textbf{45.6} 
&\textbf{44.4} 
&\textbf{38.3} 
&\textbf{46.3} 
&\textbf{12.6} 
&\textbf{11.2} 
\\
&$\beta_\textrm{s}$
&\textbf{10.5}
&\textbf{9.1} 
&\textbf{9.3} 
&\textbf{9.5} 
&\textbf{8.1} 
&\textbf{9.5} 
&\textbf{2.0} 
&-
\\
\\
CMB
&$A_\textrm{CMB}$
&\textbf{2.0} 
&\textbf{1.5} 
&\textbf{1.8} 
&\textbf{1.6} 
&\textbf{1.6} 
&\textbf{1.6} 
&1.0 
&1.0 
\\
\\
\multirow{3}{*}{Thermal dust}
&$A_\textrm{d}$
&\textbf{1.1} 
&1.0 
&1.0 
&1.0 
&1.0 
&1.0 
&1.0 
&1.0 
\\
&$\beta_\textrm{d}$ 
&\textbf{1.6} 
&\textbf{1.1} 
&\textbf{1.2} 
&\textbf{1.1} 
&1.0 
&\textbf{1.1} 
&1.0 
&-
\\
&$T_\textrm{d}$
&\textbf{1.8} 
&\textbf{1.1} 
&1.0 
&1.0 
&1.0 
&1.0 
&1.0 
&-
\\
\hline
\multicolumn{10}{c}{Polarization, curvature free to vary}\\
\hline
\multirow{3}{*}{Synchrotron}
&$A_\textrm{s}$
&\textbf{120.3} 
&\textbf{51.9} 
&\textbf{31.2} 
&\textbf{32.5} 
&\textbf{22.8} 
&\textbf{129.3}
&\textbf{4.8} 
&\textbf{3.5} 
\\
&$\beta_\textrm{s}$ 
&\textbf{3.5} 
&\textbf{2.4} 
&\textbf{2.2} 
&\textbf{2.2} 
&\textbf{2.2} 
&\textbf{5.4} 
&\textbf{1.8} 
&-
\\
&$C_\textrm{s}$
&\textbf{1.7} 
&1.0 
&0.8 
&0.9 
&0.8 
&\textbf{2.7} 
&0.8 
&-
\\
\\
CMB
&$A_\textrm{CMB}$
&1.0 
&\textbf{1.1} 
&\textbf{1.5} 
&\textbf{1.3} 
&\textbf{1.4} 
&1.0 
&0.8 
&\textbf{1.1} 
\\
\\
\multirow{3}{*}{Thermal dust}
&$A_\textrm{d}$
&1.0 
&1.0 
&1.0 
&1.0 
&1.0 
&1.0 
&1.0 
&1.0 
\\
&$\beta_\textrm{d}$ 
&1.0 
&1.0 
&\textbf{1.2} 
&\textbf{1.1} 
&1.0 
&1.0 
&0.9 
&-
\\
&$T_\textrm{d}$
&\textbf{1.3} 
&1.0 
&1.0 
&1.0 
&1.0 
&1.0 
&1.0 
&-
\\
\hline

\end{tabular}
\end{table*}

\subsection{Curved synchrotron spectrum}
\label{sec:tempCurvedSynch}
Now we consider the results when the synchrotron curvature is free to vary about the true value of zero.
The ratios of the total error volumes without C-BASS to with C-BASS are listed in Table~\ref{tab:ratioCovDet}.
The average improvement in the error volumes by the inclusion of the C-BASS data point is similar, regardless of whether the spectral curvature was free to vary or fixed, but this average hides significant variation amongst the pixels.

\begin{figure}
\centering
\includegraphics[width=\columnwidth]{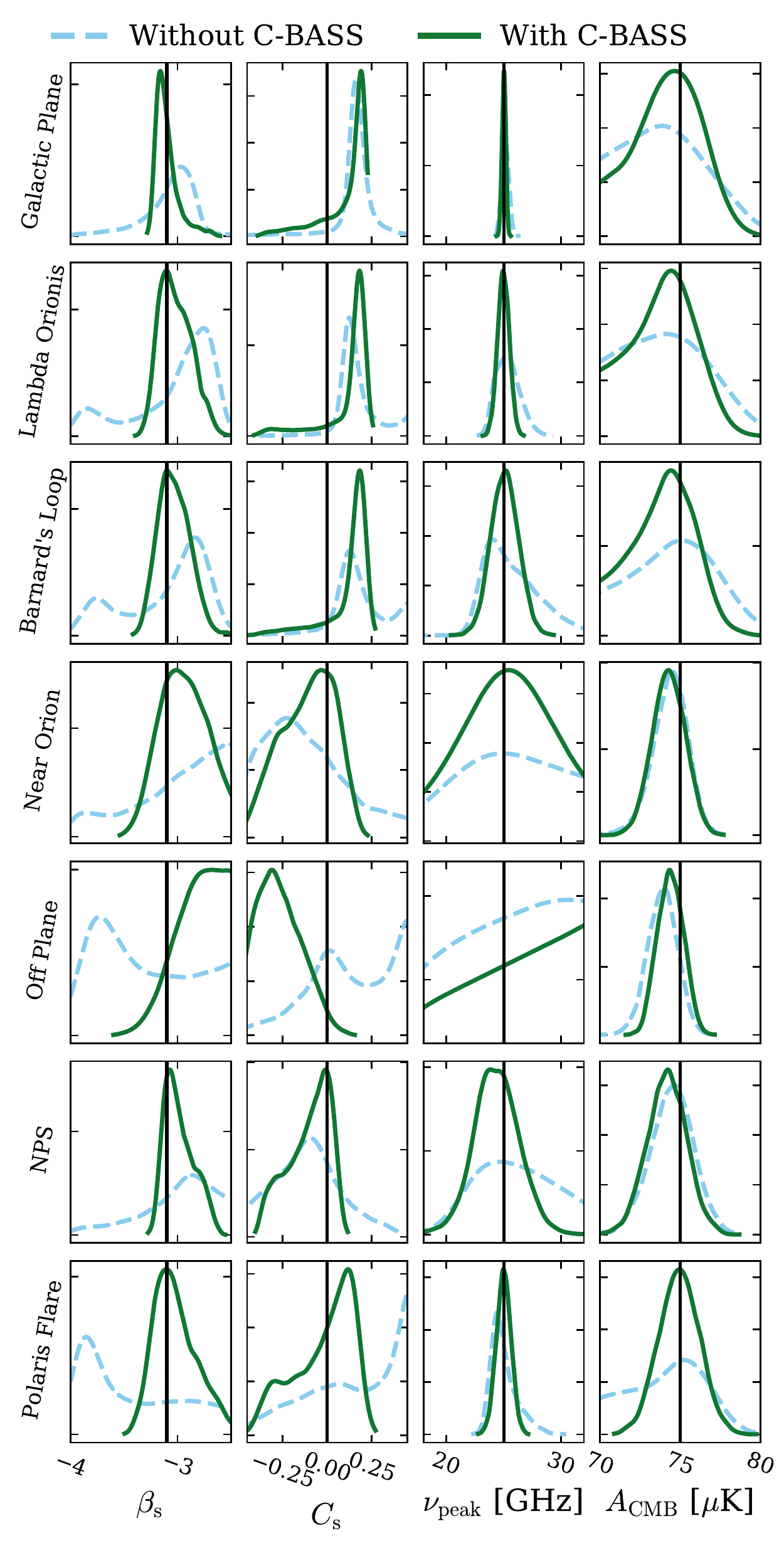}
\caption{Marginalised PDFs of the total intensity low-frequency spectral parameters and the CMB amplitude that were obtained when fitting the model to the data without C-BASS (\textit{dashed cyan}) and with C-BASS (\textit{solid green}) when the synchrotron spectral curvature is free to vary about its true value of zero.}
\label{fig:pdf_I_lowFreq_curved}
\end{figure}

\begin{figure*}
\centering
\includegraphics[width=0.2\columnwidth]{Figures/legend_2.pdf}
\includegraphics[width=\textwidth]{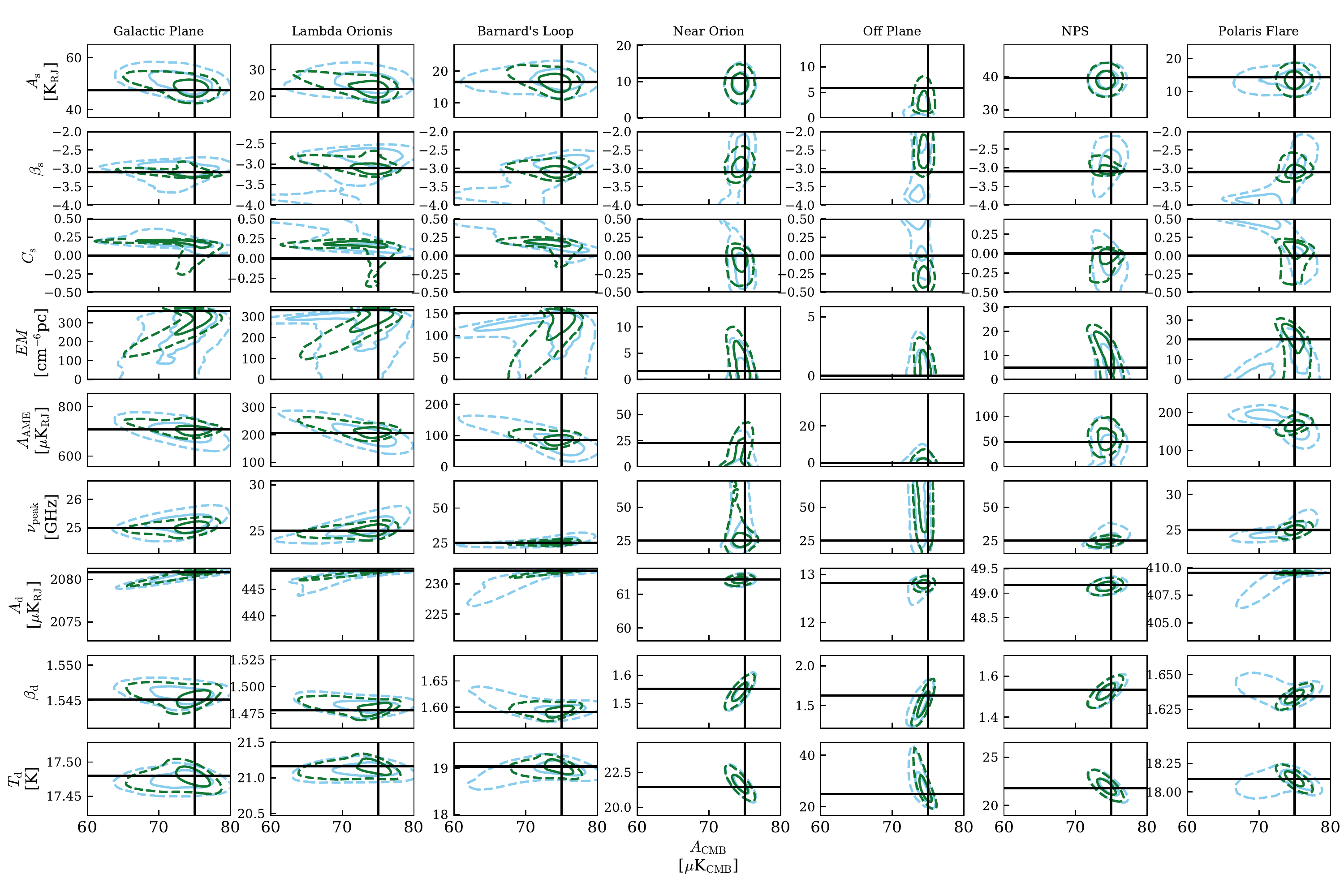}
\caption{The covariance between the CMB amplitude an the other free parameters in the total intensity pixels when the synchrotron spectral curvature is free to vary about its true value of zero in the fitting.
The contours show the 1- and 2-$\sigma$ levels in \textit{solid lines} and \textit{dashed lines} respectively.
The \textit{cyan lines} are from the chain without C-BASS.
The \textit{green lines} are from the chain with C-BASS.
The \textit{black lines} are the true parameter values.
}
\label{fig:I_curved_allCmbCov}
\end{figure*}

Marginalised PDFs for the synchrotron spectral index, spectral curvature, AME peak frequency and CMB amplitude are shown in Figure~\ref{fig:pdf_I_lowFreq_curved} and 
the improvement factors for all parameters are listed in Table~\ref{tab:totErrRatios}.
The covariance between the CMB amplitude and other parameters for each pixel are shown in Figure~\ref{fig:I_curved_allCmbCov}.

\begin{figure*}
\centering
\includegraphics[width=0.2\columnwidth]{Figures/legend_2.pdf}
\includegraphics[width=\textwidth]{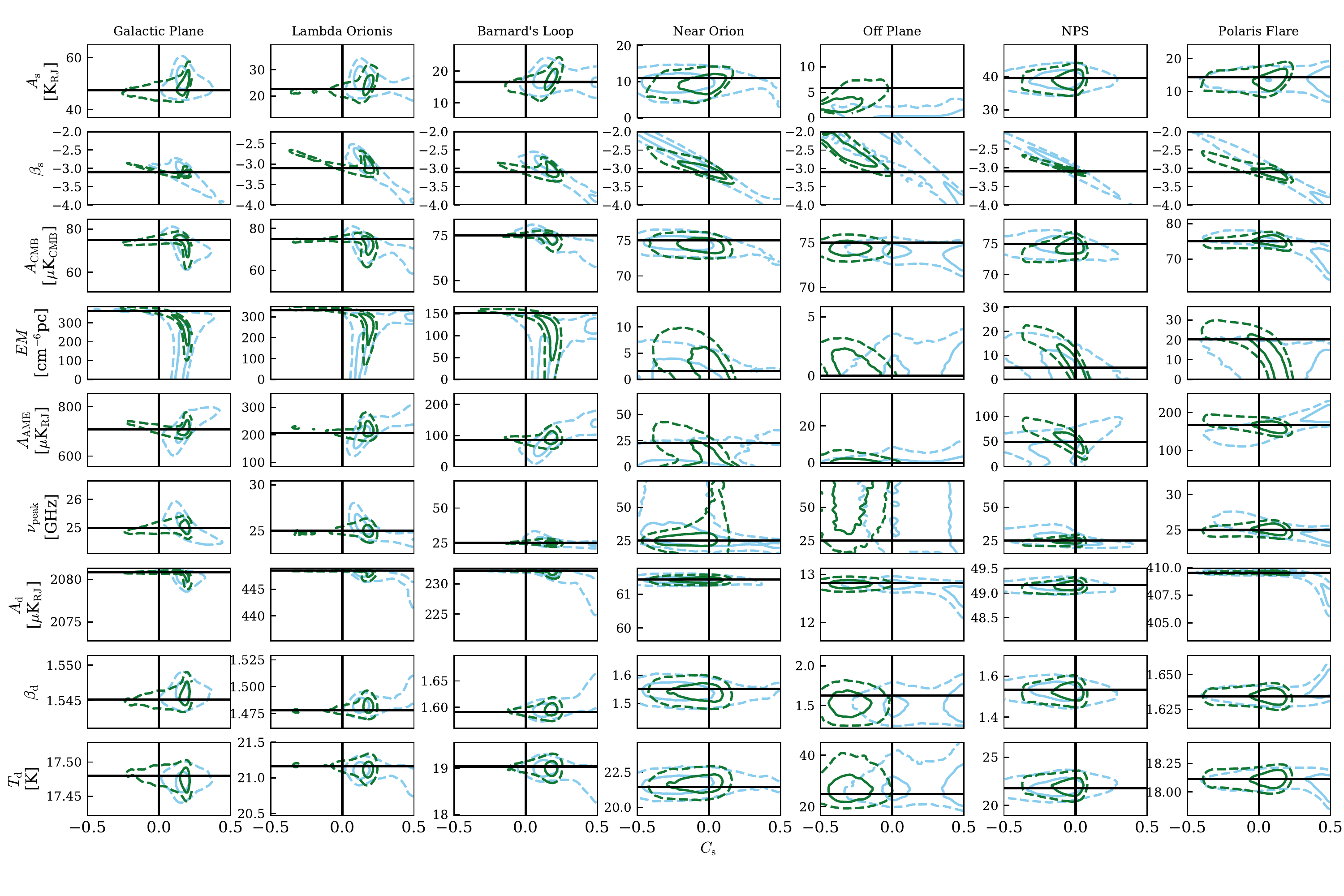}
\caption{The covariance between the synchrotron spectral curvature and the other free parameters in the total intensity pixels when the true synchrotron spectral curvature is zero.
The contours show the 1- and 2-$\sigma$ levels in \textit{solid lines} and \textit{dashed lines} respectively.
The \textit{cyan lines} are from the chain without C-BASS.
The \textit{green lines} are from the chain with C-BASS.
The \textit{black lines} are the true parameter values.
}
\label{fig:I_curved_allCsCov}
\end{figure*}

Without the C-BASS data points, the upper and lower limits on the synchrotron spectral index are determined by the prior distributions in all but the Galactic Plane pixel, the posterior distributions are often bi-modal with neither mode at the true value. 
With the C-BASS data point the spectral index parameter can be constrained in all but the Off Plane pixel and the posterior distributions peak close to the true value.
The improvement factors range from 1.1 to 2.7.
The constraints on the spectral index parameter are stronger when the curvature is fixed to its true value than when it is allowed to vary.

Allowing the synchrotron spectral curvature to vary introduces more degeneracies between the parameters and this results in non-Gaussian posterior distributions for many parameters.
These degeneracies are not completely removed by the addition of the C-BASS data point.
In high-foreground pixels, somewhat pathologically, the marginalized posterior distributions of the synchrotron spectral curvature are strongly peaked but not at the true value.
The synchrotron spectral curvature is highly degenerate with many other parameters including the CMB amplitude, free-free Emission measure, AME amplitude, synchrotron spectral index and thermal dust amplitude.
Plots of the covariance between the synchrotron spectral curvature and other parameters are shown in Figure~\ref{fig:I_curved_allCsCov}.
The degeneracies manifest as ``bananas'' in parameter covariance plots.
The resulting marginalized posterior distributions are therefore peaked away from the true value.

In every pixel with non-negligible AME amplitude, the constraint in our models on the peak frequency  is also improved by factors between 1.8 and 3.7.
The constraint on models of the CMB amplitude is improved by the addition of C-BASS in all but the NPS and Near Orion pixels.
The total error on the free-free emission measure increases after C-BASS is included in the NPS and Near Orion pixels. These two pixels have very weak free-free emission and so this worsening is not significant.
The total error volume is still decreased with the addition of C-BASS by a factor of $3\times10^2$ in the Near Orion pixel and $7\times10^4$ in the NPS pixel.

\subsection{Mis-modelling the synchrotron spectrum}
\label{sec:tempModErr}
Here we introduce a second set of simulated data, generated with a synchrotron spectral curvature of 0.15.
We fit the model to this new dataset (both with and without a C-BASS data point), firstly fixing the curvature to zero and secondly allowing the curvature to vary.
By fitting the model with a straight spectrum to the data generated with a synchrotron spectral curvature of 0.15, we introduce a modelling error.

The estimated frequency spectra for the Barnard's Loop pixel when the modelling error has been introduced are shown in the third row of Figure~\ref{fig:freqSpectraExample}.
The modelling error leads to significant underestimates of the synchrotron amplitude at higher frequencies.
However, at these higher frequencies synchrotron emission is sub-dominant to the other components and so has minimal impact on the other parameter estimates.

The total errors on the CMB amplitudes both with and without the modelling error are listed
in Table~\ref{tab:totErrs_CMB_amp}.
When the modelling error is introduced (by setting the true synchrotron curvature to 0.15 and fixing the curvature
to zero in the fitting), the total errors on the CMB amplitude parameter are similar to when there is no modelling error, particularly when the C-BASS data point is included.
Allowing the curvature to vary removes the modelling error but the extra free parameter increases the total error on the CMB amplitude.

When there is no modelling error and the synchrotron curvature is fixed to zero, without C-BASS the total errors on $A_\mathrm{CMB}$
are between 1.0 and 2.4\,$\mu$K and when C-BASS is included are between 1.0 and 1.3\,$\mu$K.

Without C-BASS, when the synchrotron spectral curvature is allowed to vary (and the true spectral curvature is still zero) the total error is around a factor of two higher compared to when the curvature is fixed.
With a free spectral curvature, including C-BASS lowers the total error and the amount depends on the level of foreground contamination in the pixel.
In low-foreground pixels, including C-BASS results in comparable total errors when the curvature is free to vary and when it is fixed at zero.
In high-foreground pixels the total error is around a factor of three greater when the curvature is free to vary compared to when it is fixed.

This demonstrates that for a synchrotron spectral curvature of 0.15, it is better with current data to accept a modelling error and fit a straight spectrum synchrotron component than to allow it to vary.

\begin{table*}
\centering
\caption{Total errors on the CMB amplitude parameter in $\mu$K for the total intensity pixels (\textit{top}) and the polarization pixels (\textit{bottom}).
The total errors were calculated from two sets of simulated data
with the true synchrotron curvature set to either 0.15 or 0.
In the fitting process the synchrotron curvature parameter was either fixed to zero or allowed to vary freely.
This introduces a modelling error in the case of simulated data with true curvature of 0.15 and when
fixing the curvature to zero in the fitting.
The columns showing the results when a modelling error has been introduced are highlighted in \textit{light grey}.}
\label{tab:totErrs_CMB_amp}
\begin{tabular}{lcrrrrcrrrr}
\hline\hline
True $C_\textrm{s}$ value && \multicolumn{4}{c}{0.15}& &\multicolumn{4}{c}{0} \\
$C_\textrm{s}$ free or fixed && \multicolumn{2}{c}{free} &  \multicolumn{2}{c}{fixed to 0}
      && \multicolumn{2}{c}{free} &  \multicolumn{2}{c}{fixed to 0}\\
\multirow{2}{*}{Data included}      && Without& With   & Without& With& & Without& With   & Without& With\\
 && C-BASS & C-BASS & C-BASS & C-BASS& & C-BASS & C-BASS & C-BASS & C-BASS\\
\hline
\multicolumn{11}{c}{Total intensity}\\
\hline
Galactic Plane   &&4.6 & 3.6  & \cellcolor{lightgray}2.7 & \cellcolor{lightgray}1.3      && 5.0 & 4.1  & 2.4 & 1.3\\
Lambda Orionis   &&4.8 & 4.1   & \cellcolor{lightgray}3.3 & \cellcolor{lightgray}1.3       && 5.7 & 4.7  & 2.3 & 1.3\\
Barnard's Loop   &&4.9  & 3.0  & \cellcolor{lightgray}3.0 & \cellcolor{lightgray}1.3      && 5.7 & 3.0  & 2.4 & 1.3\\
Near Orion       &&2.8 & 1.3   & \cellcolor{lightgray}1.3 & \cellcolor{lightgray}1.2       && 1.3 & 1.3  & 1.2 & 1.3\\
Off Plane        &&2.0 & 0.9   & \cellcolor{lightgray}1.1 & \cellcolor{lightgray}0.9       && 1.5 & 1.0  & 1.0 & 1.0\\
NPS              &&2.4 & 1.5   & \cellcolor{lightgray}2.3 & \cellcolor{lightgray}1.3       && 1.4 & 1.5  & 1.4 & 1.2\\
Polaris Flare    &&5.5 & 1.4   & \cellcolor{lightgray}2.2 & \cellcolor{lightgray}1.4       && 3.7 & 1.3  & 1.4 & 1.2\\
Zero Foregrounds &&0.5 & 0.5   & \cellcolor{lightgray}0.5 & \cellcolor{lightgray}0.5        && 0.5 & 0.5  & 0.5 & 0.5\\
\hline
\multicolumn{11}{c}{Polarization}\\
\hline
Galactic Plane  && 0.87 & 0.98 & \cellcolor{lightgray}0.65 & \cellcolor{lightgray}0.93  && 0.83 & 0.86  & 0.59 & 0.29 \\
Lambda Orionis  && 0.52 & 0.50 & \cellcolor{lightgray}0.37 & \cellcolor{lightgray}0.27  && 0.48 & 0.45  & 0.38 & 0.25 \\
Barnard's Loop  && 0.33 & 0.25 & \cellcolor{lightgray}0.25 & \cellcolor{lightgray}0.12  && 0.34 & 0.23  & 0.25 & 0.13 \\
Near Orion      && 0.24 & 0.20 & \cellcolor{lightgray}0.18 & \cellcolor{lightgray}0.10  && 0.25 & 0.18  & 0.18 & 0.11\\
Off Plane       && 0.17 & 0.14 & \cellcolor{lightgray}0.13 & \cellcolor{lightgray}0.07  && 0.18 & 0.13  & 0.13& 0.08\\
NPS             && 0.32 & 0.35 & \cellcolor{lightgray}0.22 & \cellcolor{lightgray}0.44  && 0.27 & 0.28  & 0.21 & 0.13 \\
Polaris Flare   && 0.17 & 0.20 & \cellcolor{lightgray}0.15 & \cellcolor{lightgray}0.15  && 0.17 & 0.22  & 0.15& 0.16\\
Zero Foreground && 0.05& 0.05  & \cellcolor{lightgray}0.05 & \cellcolor{lightgray}0.05  && 0.05 & 0.05  & 0.05& 0.05\\
\hline 

\end{tabular}
\end{table*}

\section{Polarization Results} 
\label{sec:PolarizationResults}

In this section we discuss the results of the parametric fits to the $3^\circ$ $B$-mode pixels.
The 5\,GHz C-BASS data point has minimal impact on the thermal dust parameters and so we focus our discussion
on the synchrotron parameters and CMB amplitude.

Table~\ref{tab:ratioCovDet} lists the ratios of the total error volumes for all eight pixels with the two sets of simulated observations (with synchrotron spectral curvature of both 0.0 and 0.15), fitting both with the spectral curvature free to vary and fixed to zero.
In all pixels the total error volume is reduced, in pixels with non-zero foreground the error volumes are reduced by factors between 50 and 2,000,000.
The improvement is greatest when the synchrotron foreground is brightest.

In Section~\ref{sec:polStraightSynch} we discuss the results where
the synchrotron spectral curvature is fixed to the true value of zero during the fitting.
In Section~\ref{sec:polCurvedSynch} we discuss the results when the curvature is allowed to vary about a true value of zero.
In Section~\ref{sec:polModErr} we discuss the impact of the modelling error, when the true curvature is 0.15 but fixing the parameter to zero in the fitting.

\subsection{Straight synchrotron spectrum}
\label{sec:polStraightSynch}

\begin{figure}
\centering
\includegraphics[width=\columnwidth]{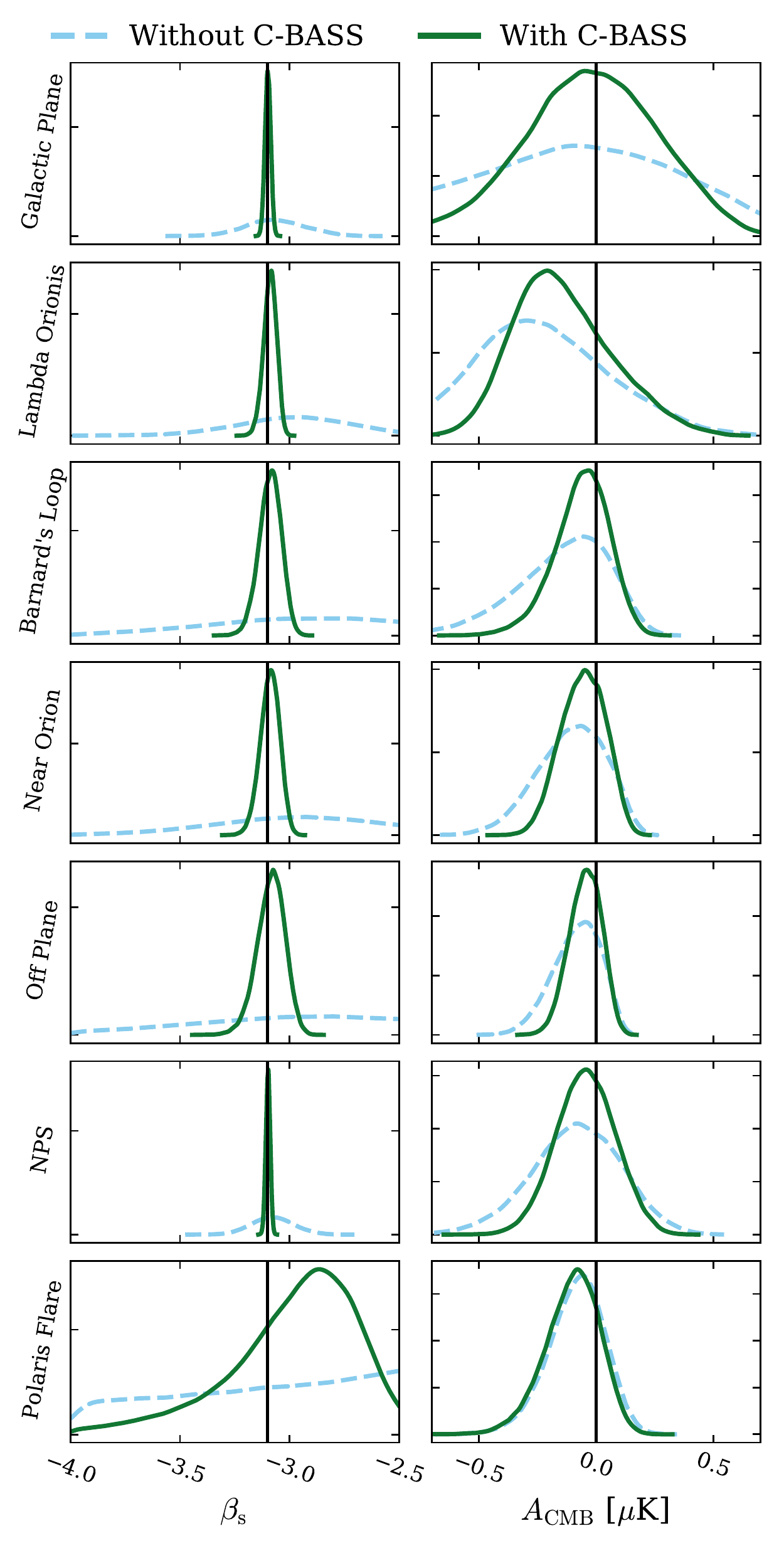}
\caption{Marginalised PDFs of the synchrotron spectral index and CMB amplitude that were obtained
without C-BASS (\textit{dashed cyan}) and with C-BASS (\textit{solid green}) in the $3^\circ$ $B-$mode pixels when the synchrotron spectral curvature was fixed to its true value of zero.}
\label{fig:pdf_0B_lowFreq_straight}
\end{figure}

Figure~\ref{fig:pdf_0B_lowFreq_straight} shows the marginalized PDFs for the synchrotron spectral index and CMB amplitude parameters in the $B$-mode polarization pixels with non-zero foreground emission.
The estimated frequency spectra for the Barnard's Loop pixel are shown in the first row of Figure~\ref{fig:freqSpectraExample}.

The \textit{Planck} and \textit{LiteBIRD} surveys  only weakly constrain estimates of the synchrotron spectral index
in the pixels with brightest synchrotron emission (Galactic Plane and NPS pixels).
Once the C-BASS data point is included, the spectral index estimate is well constrained in all but the Polaris Flare pixel, which is the pixel that has the lowest amplitude of polarized synchrotron emission.

The improvement factors on the model parameters in polarization when the synchrotron curvature is fixed to its true value 
are listed in Table~\ref{tab:totErrRatios}.

The synchrotron spectral index improvement factor is around 9--10 for all but the Polaris Flare and Off Plane pixels.
The improvement factor in the Off Plane and Polaris Flare pixels with the weakest synchrotron emission are 8.1 and 2.0 respectively.
The synchrotron amplitude in the Polaris Flare pixel is a factor of five smaller than the Off Plane pixel and a factor of ten smaller than the pixel with next lowest amplitude, Barnard's Loop.

The synchrotron amplitude at 5\,GHz is poorly constrained without the C-BASS data point.
When the C-BASS data point is included, the total error on the synchrotron amplitude is 17\,$\mu$K in the Zero Foreground pixel
and 24\,$\mu$K in all other pixels.
This is set by the thermal noise of C-BASS in $3^\circ$ pixels (24\,$\mu$K).
The improvement factors for this parameter are between 11 and 65, with the largest improvement factors in the brightest pixels.

Including the C-BASS data point improves the constraint on the CMB amplitude in all but the Polaris Flare and Zero Foregrounds pixels.
Improvement factors of up to 2 are achieved in the pixels with brightest synchrotron emission.

\subsection{Curved synchrotron spectrum}
\label{sec:polCurvedSynch}
We now consider the case of fitting a model with  $C_\textrm{s}$ free to vary when the true curvature is zero. The estimated frequency spectra for the Barnard's Loop pixel are shown in the second row of Figure~\ref{fig:freqSpectraExample}.

\begin{figure}
\centering
\includegraphics[width=\columnwidth]{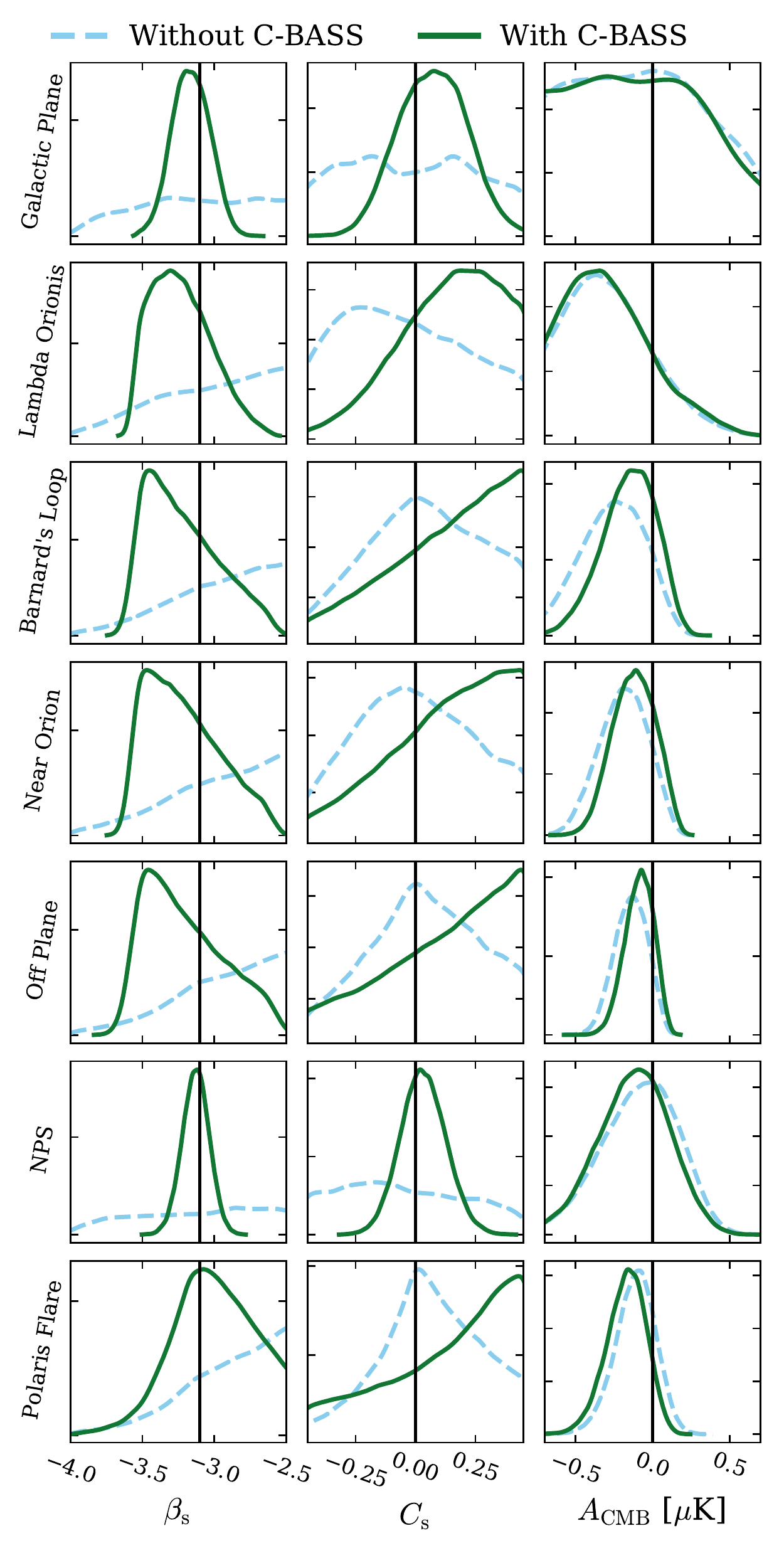}
\caption{Marginalised PDFs of the synchrotron spectral index, spectral curvature and CMB amplitude that were obtained
without C-BASS (\textit{dashed cyan}) and with C-BASS (\textit{solid green}) in the $3^\circ$ $B-$mode pixels when the synchrotron spectral curvature was free to vary.}
\label{fig:pdf_0B_lowFreq_curved}
\end{figure}

Figure~\ref{fig:pdf_0B_lowFreq_curved} shows the marginalized PDFs for the synchrotron spectral index, spectral curvature and CMB amplitude parameters for the $B$-mode polarization pixels with non-zero foreground emission. 
Neither the spectral index nor the curvature estimates can be constrained in any pixel without the C-BASS data point (the synchrotron spectral curvature posterior distributions without C-BASS are dominated by the prior distribution, which peaks at $C_\mathrm{s}=0$).
With the C-BASS data point, the spectral index and curvature estimates are only weakly constrained in the two pixels with brightest synchrotron emission (Galactic Plane and NPS pixels).
This shows that the \textit{Planck}, \textit{LiteBIRD} and C-BASS observations are not enough to fully constrain models of the synchrotron spectral curvature and more observations are needed between 5--30\,GHz.

\begin{figure}
\centering
\includegraphics[width=0.35\columnwidth]{Figures/legend_2.pdf}\\
\includegraphics[width=0.47\columnwidth]{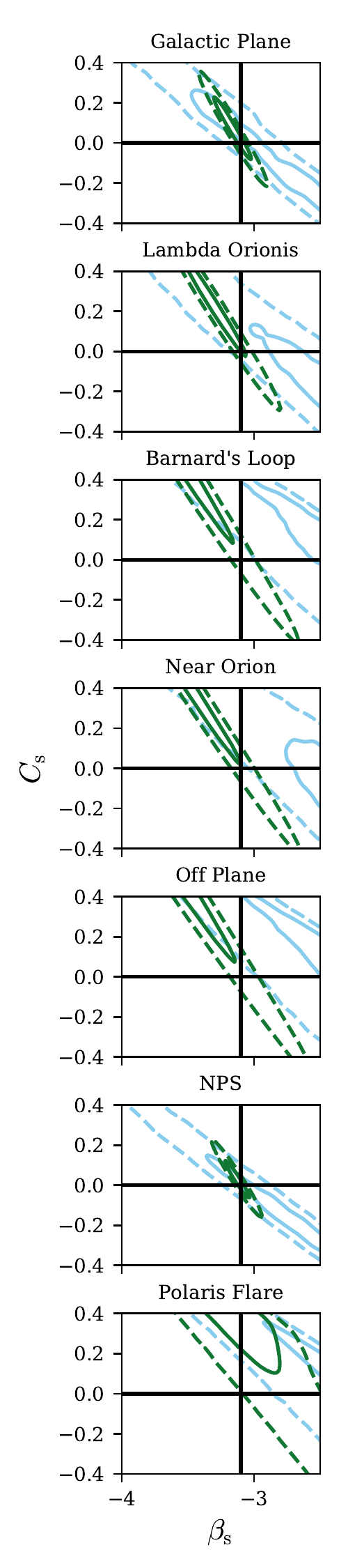}
\includegraphics[width=0.47\columnwidth]{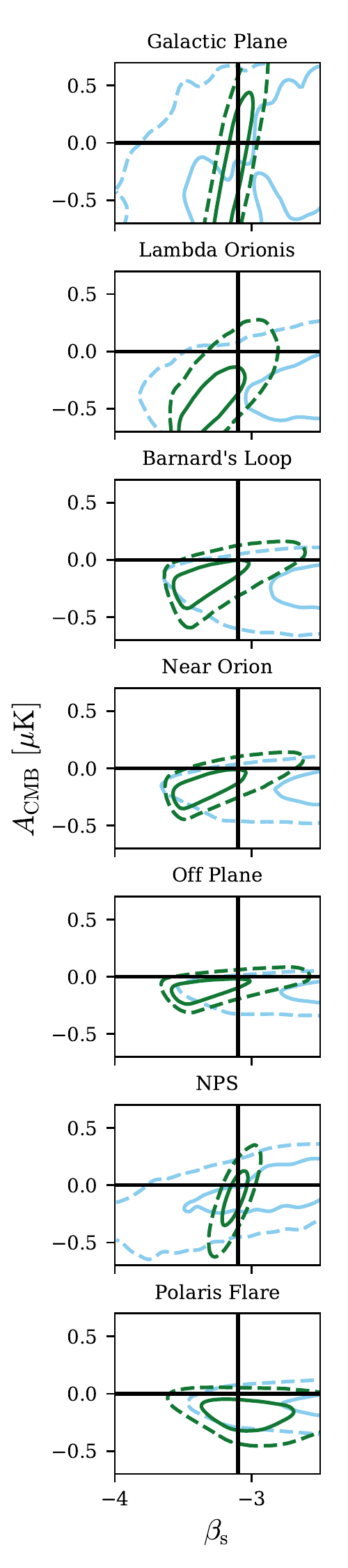}
\caption{Parameter covariances in the $3^\circ$ $B$-mode pixels, when the synchrotron spectral curvature was free to vary about a true value of zero, between the synchrotron spectral index: and the synchrotron spectral curvature (\textit{left}); and the CMB amplitude (\textit{right}).
The contours show the 1- and 2-$\sigma$ levels in \textit{solid} and \textit{dashed} lines respectively.
The \textit{cyan} lines are from the chain without C-BASS.
The \textit{green} lines are from the chain with C-BASS.
The black lines are the true parameter values.
}
\label{fig:lowFreqTwoParamCov_0B}
\end{figure}

The synchrotron curvature and spectral index parameters are highly degenerate,
as shown in the left-hand column of Figure~\ref{fig:lowFreqTwoParamCov_0B}. Shallow synchrotron emission with more negative curvature can fit the data as well
as steeper synchrotron emission with more positive curvature.

The improvement factors on the model parameters for the case of free synchrotron spectral curvature and a true value of zero 
are listed in Table~\ref{tab:totErrRatios}
The improvement on the spectral index parameter ranges between 1.8 and 5.4.
The constraint on the CMB amplitude is not improved in the Galactic Plane, NPS and Polaris Flare pixels even though the total error volumes are improved (Table~\ref{tab:ratioCovDet}).

\subsection{Mis-modelling the synchrotron spectrum}
\label{sec:polModErr}
Here we introduce a second set of simulated data, generated with a synchrotron spectral curvature of 0.15.
We fit the model to this new dataset (both with and without a C-BASS data point), firstly fixing the curvature to zero and secondly allowing the curvature to vary.
By fitting the model with a straight spectrum to the data generated with a synchrotron spectral curvature of 0.15, we introduce a modelling error.

The estimated frequency spectra for the Barnard's Loop pixel when the modelling error has been introduced are shown in the third row of Figure~\ref{fig:freqSpectraExample}.
The total errors on the CMB amplitude parameter for all eight $B$-mode pixels, for both sets of simulated data ($C_\mathrm{s}=0.0$ and 0.15), with both fixed and free spectral curvature in the fitting, and both with and without the C-BASS data point 
are listed in Table~\ref{tab:totErrs_CMB_amp}.

The modelling error only has an impact on the CMB amplitude total errors in the two pixels with brightest synchrotron emission (the Galactic Plane and NPS pixels).
In these pixels, the excess emission caused by curved synchrotron emission at higher frequencies is mis-attributed to a shallower synchrotron spectrum, excess CMB amplitude and steeper thermal dust spectrum.
In the other pixels (with fainter synchrotron emission) the modelling error has a negligible impact on the posterior estimates of the CMB amplitude parameter.

If the synchrotron spectral curvature is significant, then further low-frequency surveys will be required to constrain the CMB amplitude parameter to the levels required to detect the primordial CMB $B$-mode signal.

\section{Conclusions}
\label{sec:Conclusions}

We have
simulated the parametric fitting method of separating diffuse Galactic foregrounds from the CMB in eight pixels and
determined the additional constraining power of the 5\,GHz C-BASS data points.

In total intensity, we included simulated data from the Haslam, \textit{WMAP} and \textit{Planck} surveys.
The parametric model had ten parameters.
In summary we found:
\begin{itemize}
    \item When there was no modelling error the total error volumes were reduced by factors between 300--30,000,000 by the additional C-BASS data.
    \item When mis-modelling the synchrotron spectral curvature the total error volumes were still reduced by the addition of C-BASS, but by smaller factors of  5--20,000.
    \item The synchrotron spectral index was only convincingly constrained when C-BASS data was included.
    \item In pixels with non-negligible AME, the total error on the peak frequency parameter was reduced by factors up to 3.6 by the inclusion of C-BASS.
    \item When the synchrotron spectral curvature was fixed, including C-BASS reduced the degeneracies between parameters.
    \item When the synchrotron spectral curvature was free to vary the large degeneracies between parameters remained even with the additional C-BASS data. This shows that more low-frequency data (10--30\,GHz) are needed to constrain synchrotron spectral curvature in total intensity.
    \item Because the spectral curvature was so poorly constrained, the total error on the CMB amplitude parameter was smaller when fixing the curvature to an incorrect value than when allowing it to vary.
\end{itemize}

In polarization, we included simulated data from \textit{Planck} and the proposed \textit{LiteBIRD} satellite.
The parametric model had seven free parameters.
In summary we found:
\begin{itemize}
    \item When there was no modelling error the total error volumes were reduced by factors between 50--2,000,000.
    \item When mis-modelling the synchrotron spectral curvature, the total error volume was still reduced by the addition of C-BASS, but by smaller factors between 600--60,000.
    \item Without C-BASS, the synchrotron spectral index could only be weakly constrained in the two pixels with brightest polarized synchrotron emission. With C-BASS the spectral index was well constrained in all pixels.
    \item In pixels with the worst foreground contamination, the total error on the CMB amplitude was typically improved by factors between 1.5--2 with the inclusion of the C-BASS data.
    \item Allowing the synchrotron spectral curvature to vary introduced large degeneracies between parameters that could not be removed by the C-BASS data point.
    \item The synchrotron spectral curvature could only be weakly constrained with C-BASS in the two pixels with brightest polarized synchrotron emission.
    \item As in the total intensity case, the total errors on the CMB amplitude were smaller accepting a modelling error on the synchrotron spectrum than when allowing the spectral curvature to vary.
\end{itemize}

In summary, in total intensity the C-BASS data enables tighter constraints to be placed on low-frequency spectral parameters. In polarization, to estimate the CMB $B$-mode amplitude using pixel-by-pixel parametric fitting requires a low-frequency data point such as C-BASS.
If the synchrotron spectral curvature is believed to be significant then additional low-frequency observations (10--30\,GHz) will be needed.
Any detection of the primordial CMB $B$-mode signal would need to be tested against foreground templates (such as C-BASS) and be confirmed using multiple, independent component separation methods.,

\section*{Acknowledgements}
The \mbox{C-BASS} project is a collaboration between Oxford and Manchester Universities in the UK, the California Institute of Technology in the U.S.A., Rhodes University, UKZN and the South African Radio Observatory in South Africa, and the King Abdulaziz City for Science and Technology (KACST) in Saudi Arabia. It has been supported by the NSF awards AST-0607857, AST-1010024, AST-1212217, and AST-1616227, and NASA award NNX15AF06G, the University of Oxford, the Royal Society, STFC, and the other participating institutions. This research was also supported by the South African Radio Astronomy Observatory, which is a facility of the National Research Foundation, an agency of the Department of Science and Technology. 
CD and SH acknowledge support from an STFC Consolidated Grant (ST/P000649/1).
CD acknowledges support from an ERC Starting (Consolidator) Grant (no.~307209).
MWP acknowledges funding from a FAPESP Young Investigator fellowship, grant 2015/19936-1. We made use of the \textsc{Python} \textsc{matplotlib}, \textsc{numpy}, \textsc{healpy} \citep[a python wrapper for the  \textsc{HEALPix} package,][]{Gorski2005}, \textsc{scipy} and \textsc{PyMC} packages. (\url{http://cbass.web.ox.ac.uk}).




\bibliographystyle{mnras}
\bibliography{library} 

\begin{thebibliography}{}
\makeatletter
\relax
\def\mn@urlcharsother{\let\do\@makeother \do\$\do\&\do\#\do\^\do\_\do\%\do\~}
\def\mn@doi{\begingroup\mn@urlcharsother \@ifnextchar [ {\mn@doi@}
  {\mn@doi@[]}}
\def\mn@doi@[#1]#2{\def\@tempa{#1}\ifx\@tempa\@empty \href
  {http://dx.doi.org/#2} {doi:#2}\else \href {http://dx.doi.org/#2} {#1}\fi
  \endgroup}
\def\mn@eprint#1#2{\mn@eprint@#1:#2::\@nil}
\def\mn@eprint@arXiv#1{\href {http://arxiv.org/abs/#1} {{\tt arXiv:#1}}}
\def\mn@eprint@dblp#1{\href {http://dblp.uni-trier.de/rec/bibtex/#1.xml}
  {dblp:#1}}
\def\mn@eprint@#1:#2:#3:#4\@nil{\def\@tempa {#1}\def\@tempb {#2}\def\@tempc
  {#3}\ifx \@tempc \@empty \let \@tempc \@tempb \let \@tempb \@tempa \fi \ifx
  \@tempb \@empty \def\@tempb {arXiv}\fi \@ifundefined
  {mn@eprint@\@tempb}{\@tempb:\@tempc}{\expandafter \expandafter \csname
  mn@eprint@\@tempb\endcsname \expandafter{\@tempc}}}

\bibitem[\protect\citeauthoryear{Ali-Ha{\"{i}}moud, Hirata  \&
  Dickinson}{Ali-Ha{\"{i}}moud et~al.}{2009}]{Ali-Haimoud2008}
Ali-Ha{\"{i}}moud Y.,  Hirata C.~M.,   Dickinson C.,  2009, \mn@doi [Monthly
  Notices of the Royal Astronomical Society]
  {10.1111/j.1365-2966.2009.14599.x}, 395, 1055

\bibitem[\protect\citeauthoryear{{BICEP 2 Collaboration} et~al.,}{{BICEP 2
  Collaboration} et~al.}{2015}]{BICEP2Collaboration2015}
{BICEP 2 Collaboration} et~al., 2015, \mn@doi [Physical Review Letters]
  {10.1103/PhysRevLett.116.031302}, 116, 031302

\bibitem[\protect\citeauthoryear{Bayes \& Price}{Bayes \&
  Price}{1763}]{Bayes1763}
Bayes M.,  Price M.,  1763, \mn@doi [Philosophical Transactions of the Royal
  Society of London] {10.1098/rstl.1763.0053}, 53, 370

\bibitem[\protect\citeauthoryear{Bennett et~al.,}{Bennett
  et~al.}{2013}]{Bennett2013}
Bennett C.~L.,  et~al., 2013, \mn@doi [The Astrophysical Journal Supplement
  Series] {10.1088/0067-0049/208/2/20}, 208, 20

\bibitem[\protect\citeauthoryear{Bernardo}{Bernardo}{1979}]{Bernardo1979}
Bernardo J.~M.,  1979, {Reference Posterior Distributions for Bayesian
  Inference}, \mn@doi{10.2307/2985028}, \url
  {https://www.jstor.org/stable/2985028}

\bibitem[\protect\citeauthoryear{Carretti et~al.,}{Carretti
  et~al.}{2019}]{Carretti2019}
Carretti E.,  et~al., 2019, preprint (\mn@eprint {arXiv} {1903.09420})

\bibitem[\protect\citeauthoryear{Chluba, Hill  \& Abitbol}{Chluba
  et~al.}{2017}]{Chluba2017}
Chluba J.,  Hill J.~C.,   Abitbol M.~H.,  2017, \mn@doi [Monthly Notices of the
  Royal Astronomical Society] {10.1093/mnras/stx1982}, 472, 1195

\bibitem[\protect\citeauthoryear{Davies, Dickinson, Banday, Jaffe, G{\'{o}}rski
   \& Davis}{Davies et~al.}{2006}]{Davies2006}
Davies R.~D.,  Dickinson C.,  Banday A.~J.,  Jaffe T.~R.,  G{\'{o}}rski K.~M.,
   Davis R.~J.,  2006, \mn@doi [Monthly Notices of the Royal Astronomical
  Society] {10.1111/j.1365-2966.2006.10572.x}, 370, 1125

\bibitem[\protect\citeauthoryear{Dickinson, Davies  \& Davis}{Dickinson
  et~al.}{2003}]{Dickinson2003}
Dickinson C.,  Davies R.~D.,   Davis R.~J.,  2003, \mn@doi [Monthly Notices of
  the Royal Astronomical Society] {10.1046/j.1365-8711.2003.06439.x}, 341, 369

\bibitem[\protect\citeauthoryear{Dickinson et~al.,}{Dickinson
  et~al.}{2018}]{Dickinson2018a}
Dickinson C.,  et~al., 2018, \mn@doi [New Astronomy Reviews]
  {10.1016/j.newar.2018.02.001}, 80, 1

\bibitem[\protect\citeauthoryear{Draine}{Draine}{2011}]{Draine2011}
Draine B.~T.,  2011, {Physics of the interstellar and intergalactic medium}.
Princeton University Press

\bibitem[\protect\citeauthoryear{Draine \& Hensley}{Draine \&
  Hensley}{2016}]{Draine2016}
Draine B.~T.,  Hensley B.~S.,  2016, \mn@doi [The Astrophysical Journal]
  {10.3847/0004-637X/831/1/59}, 831, 59

\bibitem[\protect\citeauthoryear{Draine \& Lazarian}{Draine \&
  Lazarian}{1998}]{Draine1997}
Draine B.~T.,  Lazarian A.,  1998, \mn@doi [The Astrophysical Journal]
  {10.1086/311167}, 494, L19

\bibitem[\protect\citeauthoryear{Draine \& Lazarian}{Draine \&
  Lazarian}{1999}]{Draine1998a}
Draine B.~T.,  Lazarian A.,  1999, \mn@doi [The Astrophysical Journal]
  {10.1086/306809}, 512, 740

\bibitem[\protect\citeauthoryear{Dunkley et~al.,}{Dunkley
  et~al.}{2009}]{Dunkley2009}
Dunkley J.,  et~al., 2009, in AIP Conference Proceedings. AIP, pp 222--264
  (\mn@eprint {arXiv} {0811.3915}), \mn@doi{10.1063/1.3160888}, \url
  {http://arxiv.org/abs/0811.3915 http://dx.doi.org/10.1063/1.3160888
  http://aip.scitation.org/doi/abs/10.1063/1.3160888}

\bibitem[\protect\citeauthoryear{Finkbeiner}{Finkbeiner}{2003}]{Finkbeiner2003}
Finkbeiner D.~P.,  2003, \mn@doi [The Astrophysical Journal Supplement Series]
  {10.1086/374411}, 146, 407

\bibitem[\protect\citeauthoryear{Fixsen}{Fixsen}{2009}]{Fixsen2009}
Fixsen D.~J.,  2009, \mn@doi [The Astrophysical Journal]
  {10.1088/0004-637X/707/2/916}, 707, 916

\bibitem[\protect\citeauthoryear{G{\'{e}}nova-Santos
  et~al.,}{G{\'{e}}nova-Santos et~al.}{2015}]{Genova-Santos2015a}
G{\'{e}}nova-Santos R.,  et~al., 2015, in Highlights of Spanish Astrophysics
  VIII. pp 207--212, \url {http://adsabs.harvard.edu/abs/2015hsa8.conf..207G}

\bibitem[\protect\citeauthoryear{Geweke}{Geweke}{1992}]{Geweke1992}
Geweke J.,  1992, in Bernardo J.~M.,  Berger J.~O.,  Dawid A.~P.,   Smith A.
  F.~M.,  eds, Bayesian Statistics 4. Oxford University Press, pp 169--193,
  \url
  {https://pdfs.semanticscholar.org/2e86/50b01dd557ffb15113c795536ea7c6ab1088.pdf}

\bibitem[\protect\citeauthoryear{Gold et~al.,}{Gold et~al.}{2009}]{Gold:2008kp}
Gold B.,  et~al., 2009, \mn@doi [The Astrophysical Journal Supplement Series]
  {10.1088/0067-0049/180/2/265}, 180, 265

\bibitem[\protect\citeauthoryear{Gorski, Hivon, Banday, Wandelt, Hansen,
  Reinecke  \& Bartelmann}{Gorski et~al.}{2005}]{Gorski2005}
Gorski K.~M.,  Hivon E.,  Banday A.~J.,  Wandelt B.~D.,  Hansen F.~K.,
  Reinecke M.,   Bartelmann M.,  2005, \mn@doi [The Astrophysical Journal]
  {10.1086/427976}, 622, 759

\bibitem[\protect\citeauthoryear{Guzm{\'{a}}n, May, Alvarez  \&
  Maeda}{Guzm{\'{a}}n et~al.}{2011}]{Guzman2011}
Guzm{\'{a}}n A.~E.,  May J.,  Alvarez H.,   Maeda K.,  2011, \mn@doi [Astronomy
  {\&} Astrophysics] {10.1051/0004-6361/200913628}, 525, A138

\bibitem[\protect\citeauthoryear{Haslam, Salter, Stoffel  \& Wilson}{Haslam
  et~al.}{1982}]{Haslam1982}
Haslam C. G.~T.,  Salter C.~J.,  Stoffel H.,   Wilson W.~E.,  1982, Astronomy
  {\&} Astrophysics, Supplement, 47, 1

\bibitem[\protect\citeauthoryear{Hastings}{Hastings}{1970}]{Hastings1970}
Hastings W.~K.,  1970, \mn@doi [Biometrika] {10.1093/biomet/57.1.97}, 57, 97

\bibitem[\protect\citeauthoryear{Hensley \& Bull}{Hensley \&
  Bull}{2018}]{Hensley2018}
Hensley B.~S.,  Bull P.,  2018, \mn@doi [The Astrophysical Journal]
  {10.3847/1538-4357/aaa489}, 853, 127

\bibitem[\protect\citeauthoryear{Hoffman \& Gelman}{Hoffman \&
  Gelman}{2011}]{Hoffman2011}
Hoffman M.~D.,  Gelman A.,  2011, preprint (\mn@eprint {arXiv} {1111.4246})

\bibitem[\protect\citeauthoryear{Jeffreys}{Jeffreys}{1939}]{Jeffreys1939}
Jeffreys H.,  1939, {Theory of probability}, 1st edn.
Clarendon Press, Oxford

\bibitem[\protect\citeauthoryear{Jeffreys}{Jeffreys}{1946}]{Jeffreys1946}
Jeffreys H.,  1946, \mn@doi [Proceedings of the Royal Society A: Mathematical,
  Physical and Engineering Sciences] {10.1098/rspa.1946.0056}, 186, 453

\bibitem[\protect\citeauthoryear{Jew}{Jew}{2017}]{Jew2017}
Jew L.,  2017, PhD thesis, University of Oxford, \url
  {https://ora.ox.ac.uk/objects/uuid:31f0227a-84be-421a-ae46-eebe9f422767}

\bibitem[\protect\citeauthoryear{Jonas, Baart  \& Nicolson}{Jonas
  et~al.}{1998}]{Jonas1998}
Jonas J.~L.,  Baart E.~E.,   Nicolson G.~D.,  1998, \mn@doi [Monthly Notices of
  the Royal Astronomical Society] {10.1046/j.1365-8711.1998.01367.x}, 297, 977

\bibitem[\protect\citeauthoryear{Jones et~al.,}{Jones et~al.}{2018}]{Jones2018}
Jones M.~E.,  et~al., 2018, \mn@doi [Monthly Notices of the Royal Astronomical
  Society] {10.1093/mnras/sty1956}, 480, 3224

\bibitem[\protect\citeauthoryear{Keating, Timbie, Polnarev  \&
  Steinberger}{Keating et~al.}{1998}]{Keating1998}
Keating B.,  Timbie P.,  Polnarev A.,   Steinberger J.,  1998, \mn@doi [The
  Astrophysical Journal] {10.1086/305312}, 495, 580

\bibitem[\protect\citeauthoryear{LaPlace}{LaPlace}{1814}]{LaPlace1814}
LaPlace P.~S.,  1814, {Essai philosophique sur les probabilit{\'{e}}s}.
Courcier (Paris), \url {https://eudml.org/doc/203193}

\bibitem[\protect\citeauthoryear{Lawson, Mayer, Osborne  \& Parkinson}{Lawson
  et~al.}{1987}]{Lawson1987}
Lawson K.~D.,  Mayer C.~J.,  Osborne J.~L.,   Parkinson M.~L.,  1987, \mn@doi
  [Monthly Notices of the Royal Astronomical Society]
  {10.1093/mnras/225.2.307}, 225, 307

\bibitem[\protect\citeauthoryear{Liu, Creswell  \& Naselsky}{Liu
  et~al.}{2018}]{Liu2018}
Liu H.,  Creswell J.,   Naselsky P.,  2018, \mn@doi [Journal of Cosmology and
  Astroparticle Physics] {10.1088/1475-7516/2018/05/059}, 2018, 059

\bibitem[\protect\citeauthoryear{Macellari, Pierpaoli, Dickinson  \&
  Vaillancourt}{Macellari et~al.}{2011}]{Macellari2011a}
Macellari N.,  Pierpaoli E.,  Dickinson C.,   Vaillancourt J.~E.,  2011,
  \mn@doi [Monthly Notices of the Royal Astronomical Society]
  {10.1111/j.1365-2966.2011.19542.x}, 418, 888

\bibitem[\protect\citeauthoryear{Metropolis, Rosenbluth, Rosenbluth, Teller  \&
  Teller}{Metropolis et~al.}{1953}]{Metropolis1953}
Metropolis N.,  Rosenbluth A.~W.,  Rosenbluth M.~N.,  Teller A.~H.,   Teller
  E.,  1953, \mn@doi [The Journal of Chemical Physics] {10.1063/1.1699114}, 21,
  1087

\bibitem[\protect\citeauthoryear{Patil, Huard  \& Fonnesbeck}{Patil
  et~al.}{2010}]{Patil2010}
Patil A.,  Huard D.,   Fonnesbeck C.,  2010, \mn@doi [Journal of Statistical
  Software] {10.18637/jss.v035.i04}, 35, 1

\bibitem[\protect\citeauthoryear{{Planck Collaboration} et~al.,}{{Planck
  Collaboration} et~al.}{2014a}]{PlanckCollaboration2014a}
{Planck Collaboration} et~al., 2014a, \mn@doi [Astronomy {\&} Astrophysics]
  {10.1051/0004-6361/201321580}, 571, A12

\bibitem[\protect\citeauthoryear{{Planck Collaboration} et~al.,}{{Planck
  Collaboration} et~al.}{2014b}]{PlanckCollaboration2014b}
{Planck Collaboration} et~al., 2014b, \mn@doi [Astronomy {\&} Astrophysics]
  {10.1051/0004-6361/201424434}, 580, A13

\bibitem[\protect\citeauthoryear{{Planck Collaboration} et~al.,}{{Planck
  Collaboration} et~al.}{2014c}]{PlanckCollaboration2014}
{Planck Collaboration} et~al., 2014c, \mn@doi [Astronomy {\&} Astrophysics]
  {10.1051/0004-6361/201425034}, 586, A133

\bibitem[\protect\citeauthoryear{{Planck Collaboration} et~al.,}{{Planck
  Collaboration} et~al.}{2015a}]{Ade2015}
{Planck Collaboration} et~al., 2015a, \mn@doi [Astronomy {\&} Astrophysics]
  {10.1051/0004-6361/201424082}, 576, A104

\bibitem[\protect\citeauthoryear{{Planck Collaboration} et~al.,}{{Planck
  Collaboration} et~al.}{2015b}]{PlanckCollaboration2015}
{Planck Collaboration} et~al., 2015b, \mn@doi [Astronomy {\&} Astrophysics]
  {10.1051/0004-6361/201525967}, 594, A10

\bibitem[\protect\citeauthoryear{{Planck Collaboration} et~al.,}{{Planck
  Collaboration} et~al.}{2015c}]{Ade2016}
{Planck Collaboration} et~al., 2015c, \mn@doi [Astronomy {\&} Astrophysics]
  {10.1051/0004-6361/201526803}, 594, A25

\bibitem[\protect\citeauthoryear{Platania, Burigana, Maino, Caserini,
  Bersanelli, Cappellini  \& Mennella}{Platania et~al.}{2003}]{Platania2003}
Platania P.,  Burigana C.,  Maino D.,  Caserini E.,  Bersanelli M.,  Cappellini
  B.,   Mennella A.,  2003, \mn@doi [Astronomy {\&} Astrophysics]
  {10.1051/0004-6361:20031125}, 410, 847

\bibitem[\protect\citeauthoryear{Raftery, Raftery  \& Lewis}{Raftery
  et~al.}{1995}]{Raftery1995}
Raftery A.~E.,  Raftery A.~E.,   Lewis S.~M.,  1995, IN PRACTICAL MARKOV CHAIN
  MONTE CARLO, pp 115----130

\bibitem[\protect\citeauthoryear{Reich \& Reich}{Reich \&
  Reich}{1988}]{Reich1988}
Reich P.,  Reich W.,  1988, Astronomy and Astrophysics Supplement Series, 74, 7

\bibitem[\protect\citeauthoryear{Remazeilles \& Chluba}{Remazeilles \&
  Chluba}{2018}]{Remazeilles2018}
Remazeilles M.,  Chluba J.,  2018, \mn@doi [Monthly Notices of the Royal
  Astronomical Society] {10.1093/mnras/sty1034}

\bibitem[\protect\citeauthoryear{Remazeilles, Dickinson, Banday, Bigot-Sazy  \&
  Ghosh}{Remazeilles et~al.}{2015}]{Remazeilles2014}
Remazeilles M.,  Dickinson C.,  Banday A.~J.,  Bigot-Sazy M.-A.,   Ghosh T.,
  2015, \mn@doi [Monthly Notices of the Royal Astronomical Society]
  {10.1093/mnras/stv1274}, 451, 4311

\bibitem[\protect\citeauthoryear{Remazeilles, Dickinson, Eriksen  \&
  Wehus}{Remazeilles et~al.}{2016}]{Remazeilles2015}
Remazeilles M.,  Dickinson C.,  Eriksen H. K.~K.,   Wehus I.~K.,  2016, \mn@doi
  [Monthly Notices of the Royal Astronomical Society] {10.1093/mnras/stw441},
  458, 2032

\bibitem[\protect\citeauthoryear{Rybicki \& Lightman}{Rybicki \&
  Lightman}{1985}]{Rybicki1985}
Rybicki G.~B.,  Lightman A.~P.,  1985, {Radiative Processes in Astrophysics}.
Wiley-VCH Verlag GmbH, Weinheim, Germany, \mn@doi{10.1002/9783527618170}, \url
  {http://doi.wiley.com/10.1002/9783527618170}

\bibitem[\protect\citeauthoryear{Silsbee, Ali-Ha{\"{i}}moud  \& Hirata}{Silsbee
  et~al.}{2011}]{Silsbee2010}
Silsbee K.,  Ali-Ha{\"{i}}moud Y.,   Hirata C.~M.,  2011, \mn@doi [Monthly
  Notices of the Royal Astronomical Society]
  {10.1111/j.1365-2966.2010.17882.x}, 411, 2750

\bibitem[\protect\citeauthoryear{Suzuki et~al.,}{Suzuki
  et~al.}{2018}]{Suzuki2018}
Suzuki A.,  et~al., 2018, \mn@doi [Journal of Low Temperature Physics]
  {10.1007/s10909-018-1947-7}, 193, 1048

\bibitem[\protect\citeauthoryear{Vidal, Dickinson, Davies  \& Leahy}{Vidal
  et~al.}{2015}]{Vidal2014}
Vidal M.,  Dickinson C.,  Davies R.~D.,   Leahy J.~P.,  2015, \mn@doi [Monthly
  Notices of the Royal Astronomical Society] {10.1093/mnras/stv1328}, 452, 656

\makeatother
\end{thebibliography}






\bsp	
\label{lastpage}
\end{document}